\documentclass[12pt,reqno]{amsart}

\usepackage{cite,color}

\newtheorem{theorem}{Theorem}[section]

\theoremstyle{definition}
\newtheorem{remark}[theorem]{Remark}

\numberwithin{equation}{section}

\newcommand{\T}{{}^{\mathrm t}}
\begin{document}

	\title[Violation of Parity and Flavor Symmetries in NJL Model]{Violation of Parity and Flavor Symmetries in a Nambu--Jona-Lasinio Model}

	\author[Y. Goto]{Yukimi Goto\textsuperscript{1}}
	\thanks{\textsuperscript{1} Gakushuin University,~Department of Mathematics, Mejiro, Toshima-ku, Tokyo 171-8588, Japan.
		Present affiliation: Graduate School of Mathematical Sciences, University of Tokyo, Komaba, Meguro-ku, Tokyo 153-8914, Japan. Email:~{\tt  yukimi@ms.u-tokyo.ac.jp}}

	\author[T. Koma]{Tohru Koma\textsuperscript{2}}
	\thanks{\textsuperscript{2} Gakushuin University (retired), Department of Physics, Mejiro, Toshima-ku, Tokyo 171-8588, Japan}

	\maketitle

	\noindent
	{\bf Abstract:}
	We study a lattice Nambu--Jona-Lasinio model with certain continuous chiral and two-flavor symmetries.
	For the Hamiltonian of the model, we construct a ground state which breaks the parity and flavor symmetries.
	In our argument, the chiral symmetry plays a crucial role in proving the violation of the parity and flavor symmetries,
	although the model does not contain the so-called Wilson term.

	\tableofcontents

	\section{Introduction}
	\label{Sec:Intro}

	In the present paper, we deal with a lattice Nambu--Jona-Lasinio model with certain continuous chiral and two-flavor symmetries.
	A model of this type in the lattice regularization with Wilson fermions is expected
	to exhibit the violation of the flavor symmetry and the parity in the ground state~\cite{ABG}.
	In lattice quantum chromodynamics (QCD) literature, such a phase structure is referred to as Aoki phase,
	going back to the pioneering works by Aoki~\cite{Aoki1,Aoki2}.
	For Wilson fermions, it was predicted in~\cite{Aoki1,Aoki2} that, in the broken phase for the flavor symmetry and the parity,
	the charged pions can be interpreted as Nambu--Goldstone (massless) bosons, despite the absence of chiral symmetry.
	This expectation was strengthened by computer simulations of QCD \cite{AG, Aoki4}.

	For the present Nambu--Jona-Lasinio model, we construct a ground state which breaks the flavor symmetry and the parity.
	More precisely, we introduce certain two observables to detect the violation of the flavor symmetry and the parity,
	and for these observables, we prove the existence of a non-vanishing spontaneous magnetization in the ground state.
	Technically, the method of fermion reflection positivity~\cite{JP} is crucial for the proof,
	as in our previous works~\cite{GK, GK2}. However, it should be noted that if we introduce a symmetry-breaking field,
	the existence of the corresponding spontaneous magnetization is always equivalent to the non-vanishing
	expectation value of the order parameter $\bar\Psi \Psi$ due to the symmetries
	of the present model. Here, we have used the standard notation
	for the order parameter $\bar\Psi \Psi=\Psi^\dagger \gamma_0\Psi$.
	In Sec.~\ref{Sect.model} below, we will give the precise definition of
	the set of the whole order parameters to detect the symmetry breaking.
	Clearly, the order parameter $\bar\Psi \Psi$ is invariant under both the flavor rotation and the parity transformation.
	Thus, the usual procedure does not work for showing the desired spontaneous symmetry breaking.
	To overcome this difficulty, we use a certain trial state based on an idea introduced by Horsch and von der Linden~\cite{HL}.
	In particular, we can show the existence of the simultaneous non-vanishing spontaneous magnetizations
	in two directions using the two order parameters.
	This approach is one of the principal novel ingredients in our paper.
	Thus, the above conjecture is proved for our model without any approximation.
	However, we have not been able to prove that there exists an infinite-volume ground state
	that is pure and breaks both flavor symmetry and parity.
	Namely, the ground state which we constructed may consist of two or more pure states,
	and each pure state may not break the two symmetries simultaneously.
	Unfortunately, we cannot deny the possibility.
	But we can expect that such a situation never occurs,
	since the two observables for the two order parameters can be simultaneously diagonalized
	in the finite volume systems by the corresponding commutativity $[\gamma_0\tau_1, i\gamma_0\gamma_5\tau_2] = 0$
	in the Hilbert space. Here, $\tau_j$ are the Pauli matrices which act on the flavor degrees of freedom.
	For further details, see Remark~\ref{rem.th} below.

	In the rest of this section, we roughly describe the present model
	and our idea to prove the violation of the parity and flavor symmetries.
	The precise mathematical definitions will be given later.
	For this purpose, we start from our motivation.
	Namely, the numerical simulations of QCD showed the violation of the parity and flavor symmetries \cite{AG}.
	The Lagrangian of QCD is given by
	\begin{equation*}
		\mathcal{L}_0=\sum_{i=1}^{N_{\rm f}}\bar{\Psi}_i\gamma_\mu(\partial^\mu+ig_0 A_\mu)\Psi_i+\frac{1}{4}F_{\mu\nu}F^{\mu\nu},
	\end{equation*}
	where $\Psi_i$ are the quark fields with $N_{\rm f}$ flavors, $A_\mu$ is the gauge field, $F_{\mu\nu}$ is its field strength,
	and $g_0$ is the coupling constant in the standard notation. Clearly, this Lagrangian $\mathcal{L}_0$ has
	the symmetry of the SU$({\rm N}_{\rm f})$ rotation about the flavors. In addition, it is invariant
	under the usual chiral transformation,
	\begin{equation}
		\Psi_j\rightarrow e^{i\theta_j\gamma_5}\Psi_j\quad \mbox{for \ } j=1,2,\cdots, N_{\rm f},
		\label{chiraltrans0}
	\end{equation}
	with the $N_{\rm f}$ angles, $\theta_j\in[0,2\pi),\ j=1,2,\ldots,N_{\rm f}$.

	Next, we go back to the original idea of Nambu \cite{Nambu} for getting the effective Lagrangian.
	In the following, we will treat only the case with two flavors, i.e., $N_{\rm f}=2$, because the generalization is straightforward.
	We also write
	\begin{equation*}
		\Psi=\begin{pmatrix}
			\Psi_1 \\ \Psi_2
		\end{pmatrix}
		=\begin{pmatrix}
			\Psi_{\rm u} \\ \Psi_{\rm d}
		\end{pmatrix}.
	\end{equation*}
	By formally integrating out the gauge fields $A_\mu$, the resulting Lagrangian $\mathcal{L}$ can be written in terms of
	only the quark fields $\Psi_i$,
	and the Lagrangian becomes the effective one for the self-interacting fermion system.
	In the following, we will not treat the color degrees of freedom.
	As an approximation for the dominant contributions in the interaction terms, the four-fermion interactions
	have been widely used so far. The explicit form of the effective Lagrangian is given by
	\begin{equation}
		\mathcal{L}=\bar{\Psi}\gamma_\mu\partial^{\mu}\Psi+\tilde{g}\Bigl\{(\bar{\Psi}\Psi)^2+(\bar{\Psi}i\gamma_5\Psi)^2
		+\sum_{j=1}^3[(\bar{\Psi}\tau_j\Psi)^2+(\bar{\Psi}i\gamma_5\tau_j\Psi)^2]\Bigr\}
		\label{LNJLmodel}
	\end{equation}
	with the effective coupling constant $\tilde{g}$, where $\tau_j$ are the Pauli matrices which act only on the flavor indices.
	From a physical viewpoint, we assume that the effective interactions between two quarks,
	which are mediated by the gauge fields $A_\mu$, are attractive.
	Therefore, we choose $\tilde{g}>0$ for the effective coupling constant.
	The above $\mathcal{L}$ is nothing but the Lagrangian of a generalized Nambu--Jona-Lasinio model \cite{NJL,NJL2,Hatsuda}.
	Clearly, it is invariant under the SU(2) flavor rotation. Further, this Lagrangian $\mathcal{L}$
	is also invariant under the above chiral transformation (\ref{chiraltrans0}). The latter invariance will be checked in
	Sec.~\ref{sec:chiralsymm} below for the corresponding Hamiltonian on a lattice.
	Since the chiral symmetry plays a crucial role in proving the violation of the parity and flavor symmetries,
	we cannot include the so-called Wilson term in the Hamiltonian. Therefore, the well-known doubling problem arises in our model.
	In the original works on Aoki phase~\cite{Aoki1,Aoki2,Aoki3,Aoki4},
	it was also argued that the expectation value of the flavor singlet pseudoscalar operator, $i\bar{\Psi} \gamma_5 \Psi$,
	must vanish as a result of the Vafa--Witten theorem~\cite{VW}.
	This means that the $\eta$ meson remains massive in the broken phase for the parity and flavor symmetries,
	thereby solving the U(1) problem on the lattice.
	However, since our model presumably lacks a desirable continuum limit
	that reproduces the chiral anomaly, due to the presence of doublers~\cite{NN1,NN2,CGJ,Nakamura},
	it may have a different physical interpretation compared to the standard picture of Aoki phase.

	Since the interactions of the Lagrangian $\mathcal{L}$ are attractive, the corresponding interaction Hamiltonian,
	whose explicit form is given by (\ref{Hint}) below on a lattice,
	looks very similar to the ferromagnetic Heisenberg spin Hamiltonian, although their internal degrees of freedom are
	totally different from each other. As is well known, in the ground states of the Heisenberg ferromagnet,
	the spins align in the same direction, and the rotational symmetry of the spins is spontaneously broken.

	In the model given by the above Lagrangian $\mathcal{L}$ of (\ref{LNJLmodel}), the observables which correspond to
	the spin components of the Heisenberg model contain two observables \cite{AG},
	$\bar{\Psi}\tau_1\Psi$ and $\bar{\Psi}i\gamma_5\tau_2\Psi$.
	In the present paper, we construct a ground state such that the ground-state expectation values for these two observables
	are simultaneously non-vanishing. This result implies the violation of the parity and flavor symmetries.
	In our proof, we need to treat the above two observables because of the symmetries of the present system.

	A reader might think that it is impossible that there appear two spontaneous magnetizations with different directions
	in a single phase because the spontaneous mass generation about the order parameter $\bar{\Psi}\tau_1\Psi$
	suppresses the long-range order of the second order parameter $\bar{\Psi}i\gamma_5\tau_2\Psi$.
	In Appendix~\ref{Sec.EH}, we give a
	plausible argument that two such spontaneous magnetizations are possible to emerge	in different directions.
	More precisely, we introduce an effective massive model. Since a single spontaneous magnetization is proved to appear
	in our mathematical argument,
	we replace the	effect of one of the two spontaneous magnetizations by the corresponding effective mass term.
	For this effective massive model,\footnote{When introducing the masses of fermions, the pseudoscalar channels of
		the effective four-fermion interactions are expected to become dominant compared to the scalar ones in the Nambu-Jona-Lasinio model
		\cite{Braghin1,Braghin2}.}
	we prove the existence of the long-range order
	for the second order parameter $\bar{\Psi}i\gamma_5\tau_2\Psi$. This implies the parity violation.
	\medskip

	The present paper is organized as follows: We give the precise definition of the Hamiltonian of the Nambu--Jona-Lasinio model
	in Sec.~\ref{Sect.model}, and the chiral symmetry of the Hamiltonian is checked in Sec.~\ref{sec:chiralsymm}.
	Our main result is given in Sec.~\ref{sec:result}. The reflection positivity of the model is proved
	in the three Sections~\ref{sec:Reflec.Preli}--\ref{sec:x2plane}.
	The Gaussian domination and the infrared bound are obtained in Sec.~\ref{GdomiInfrab}.
	The existence of the long-range order is proved in Sec.~\ref{sec:LRO}.
	Appendices~\ref{ChiralRot} and \ref{LboundcalE} are respectively devoted to technical relations and estimates.
	In Appendices~\ref{Sec.EH} and~\ref{Sec.RPEM}, we treat the above effective massive model
	that actually shows the desired parity violation.

	\section{Hamiltonian of the Nambu--Jona-Lasinio model}
	\label{Sect.model}

	In order to derive the Hamiltonian of the Nambu--Jona-Lasinio model (\ref{LNJLmodel}) on a lattice, consider first
	the case of the single-flavor. The corresponding Lagrangian of the Nambu--Jona-Lasinio model is given by \cite{NJL,NJL2,Hatsuda}
	$$
	\mathcal{L}=\bar{\Psi}\gamma_\mu\partial^{\mu}\Psi+\tilde{g}[(\bar{\Psi}\Psi)^2-(\bar{\Psi}\gamma_5\Psi)^2]
	$$
	with the effective coupling constant $\tilde{g}\ge 0$.
	The four-fermion interaction for the Dirac fermion fields $\Psi$ is formally written
	\begin{align*}
		&(\bar{\Psi}\Psi)^2-(\bar{\Psi}\gamma_5\Psi)^2\\
		&=\frac{1}{2}(\bar{\Psi}\Psi+\bar{\Psi}\gamma_5\Psi)(\bar{\Psi}\Psi-\bar{\Psi}\gamma_5\Psi)
		+\frac{1}{2}(\bar{\Psi}\Psi-\bar{\Psi}\gamma_5\Psi)(\bar{\Psi}\Psi+\bar{\Psi}\gamma_5\Psi)\\
		&=2\biggl[\bar{\Psi}\frac{1+\gamma_5}{2}\Psi\cdot \bar{\Psi}\frac{1-\gamma_5}{2}\Psi
		+\bar{\Psi}\frac{1-\gamma_5}{2}\Psi\cdot \bar{\Psi}\frac{1+\gamma_5}{2}\Psi\biggr]\\
		&=2\biggl[\Psi^\dagger\gamma_0\frac{1+\gamma_5}{2}\Psi\cdot \Psi^\dagger \gamma_0\frac{1-\gamma_5}{2}\Psi
		+\Psi^\dagger \gamma_0\frac{1-\gamma_5}{2}\Psi\cdot \Psi^\dagger \gamma_0\frac{1+\gamma_5}{2}\Psi\biggr].
	\end{align*}
	We want to define the corresponding interaction on the three-dimensional cubic lattice $\Lambda\subset\mathbb{Z}^3$
	in the Hamiltonian formalism. We simply introduce
	\begin{equation*}
		\Gamma^{(+)}(x):=\Psi^\dagger(x)\gamma_0\cdot\frac{1}{2}(1+\gamma_5)\Psi(x)
	\end{equation*}
	and
	\begin{equation*}
		\Gamma^{(-)}(x):=\Psi^\dagger(x)\gamma_0\cdot\frac{1}{2}(1-\gamma_5)\Psi(x),
	\end{equation*}
	where $\Psi(x)$ is the four-component Dirac-field operator at the site $x\in\Lambda$.
	Then, from the above observations, one can consider the nearest-neighbor interaction given by
	\begin{equation*}
		\Gamma^{(+)}(x)\Gamma^{(-)}(y)+\Gamma^{(-)}(x)\Gamma^{(+)}(y)
	\end{equation*}
	for the nearest-neighbor sites, $x,y\in\Lambda$. We also introduce
	\begin{equation*}
		\Gamma^{(1)}(x):=\Gamma^{(+)}(x)+\Gamma^{(-)}(x)
	\end{equation*}
	and
	\begin{equation*}
		\Gamma^{(2)}(x):=i[\Gamma^{(+)}(x)-\Gamma^{(-)}(x)].
	\end{equation*}
	Clearly, one has
	\begin{equation}
		\label{Gammas}
		\Gamma^{(1)}(x)=\Psi^\dagger(x)\gamma_0\Psi(x)
		\quad \mbox{and}\quad \Gamma^{(2)}(x)=\Psi^\dagger(x)i\gamma_0\gamma_5\Psi(x).
	\end{equation}
	In addition, one notices
	\begin{equation*}
		\Gamma^{(1)}(x)\Gamma^{(1)}(y)+\Gamma^{(2)}(x)\Gamma^{(2)}(y)
		=2\bigl[\Gamma^{(+)}(x)\Gamma^{(-)}(y)+\Gamma^{(-)}(x)\Gamma^{(+)}(y)\bigr].
	\end{equation*}
	From these observations, one can consider the interaction Hamiltonian,
	\begin{equation*}
		H_{\rm int}=-g\sum_{x,y:|x-y|=1}\bigl[\Gamma^{(1)}(x)\Gamma^{(1)}(y)+\Gamma^{(2)}(x)\Gamma^{(2)}(y)\bigr],
	\end{equation*}
	with the coupling constant $g>0$, i.e., we assume that the interaction is attractive, following the idea of \cite{NJL}.
	Here, the sum is over all the nearest-neighbor pairs, $x,y\in\Lambda$.
	The precise definition of the finite lattice which we consider is given by
	$$
	\Lambda := \{x = (x^{(1)}, x^{(2)}, x^{(3)}) \in \mathbb{Z}^3 \colon -L+1\le x^{(i)}\le  L,\ i=1,2,3 \}
	$$
	with a positive integer $L$, and the periodic boundary condition. The lattice $\Lambda$ can be considered as
	the three-dimensional torus. We write $e_\mu$ for the unit vector whose $\mu$-th component is $1$.

	Next, consider the two-flavor Dirac fermions.
	The four-component Dirac field $\Psi_{\rm f}(x)$ with the flavor ${\rm f}={\rm u,d}$ at the site $x\in\Lambda$ is given by
	\[
	\Psi_{\rm f}(x):=
	\begin{pmatrix}
		\psi_{\rm f}^{(1)}(x) \\ \psi_{\rm f}^{(2)}(x) \\ \psi_{\rm f}^{(3)}(x) \\ \psi_{\rm f}^{(4)}(x)
	\end{pmatrix},
	\quad
	\Psi_{\rm f}^\dagger(x):=\left(	\psi_{\rm f}^{(1)}(x)^\dagger, \psi_{\rm f}^{(2)}(x)^\dagger, \psi_{\rm f}^{(3)}(x)^\dagger, \psi_{\rm f}^{(4)}(x)^\dagger \right)
	\]
	where $\psi_{\rm f}^{(i)}(x)$ is the fermion operator for the component $i\in\{1,2,3,4\}$ at the site $x\in\Lambda$
	with the flavor ${\rm f}\in\{{\rm u,d}\}$.
	These satisfy the anti-commutation relations,
	\[
	\{\psi_{\rm f}^{(i)}(x), [\psi_{\rm f'}^{(j)}(y)]^\dagger \} = \delta_{x, y}\delta_{i,j}\delta_{\rm f,f'}, \quad
	\{\psi_{\rm f}^{(i)}(x), \psi_{\rm f'}^{(j)}(y)\}=0,
	\]
	for the sites $x, y \in \Lambda$, $i,j\in\{1,2,3,4\}$ and ${\rm f, f'}\in\{\rm u,d\}$. We write
	\begin{equation*}
		\Psi(x):=\begin{pmatrix}
			\Psi_{\rm u}(x) \\ \Psi_{\rm d}(x)
		\end{pmatrix},
		\quad
		\Psi^\dagger(x):=
		\left(\Psi^\dagger_{\rm u}(x), \Psi^\dagger_{\rm d}(x)\right)
	\end{equation*}
	for short.
	In order to simplify the notation, we will use $\mathcal{M} \Psi(x)$ to mean
	\[
	\mathcal{M}\Psi(x)=
	\begin{pmatrix}
		\mathcal{M}	\Psi_{\rm u}(x) \\ \mathcal{M}\Psi_{\rm d}(x)
	\end{pmatrix},
	\]
	for any $4\times 4$ matrix $\mathcal{M}$.
	Our Hamiltonian of the Nambu--Jona-Lasinio model~\cite{Hatsuda} is given by
	\begin{equation}
		\label{HamLam}
		H^{(\Lambda)}:=H_{\rm K}^{(\Lambda)}+H_{\rm int}^{(\Lambda)},
	\end{equation}
	where the first term is the usual kinetic Hamiltonian given by
	\begin{align}
		\label{HK}
		H_{\rm K}^{(\Lambda)}&:=i\kappa \sum_{x\in\Lambda\subset\mathbb{Z}^3}
		\Bigl\{[\Psi^\dagger(x)\alpha_1\Psi(x+e_1)-\Psi^\dagger(x+e_1)\alpha_1\Psi(x)] \nonumber\\
		& \qquad \qquad +[\Psi^\dagger(x)\alpha_2\Psi(x+e_2)-\Psi^\dagger(x+e_2)\alpha_2\Psi(x)] \nonumber\\
		& \qquad \qquad +[\Psi^\dagger(x)\alpha_3\Psi(x+e_3)-\Psi^\dagger(x+e_3)\alpha_3\Psi(x)]\Bigr\},
	\end{align}
	with the hopping parameter $\kappa\in\mathbb{R}$, where
	\begin{equation*}
		\alpha_i=\begin{pmatrix}
			0 & \sigma_i \\ \sigma_i & 0
		\end{pmatrix}
		\quad \mbox{for \ } i=1,2,3
	\end{equation*}
	with the Pauli matrices,
	\begin{equation*}
		\sigma_1=\begin{pmatrix}
			0 & 1 \\ 1 & 0
		\end{pmatrix},
		\quad
		\sigma_2=\begin{pmatrix}
			0 & -i \\ i & 0
		\end{pmatrix}
		\quad \mbox{and} \quad
		\sigma_3=\begin{pmatrix}
			1 & 0 \\ 0 & -1
		\end{pmatrix}.
	\end{equation*}
	The second term is the interaction Hamiltonian given by
	\begin{equation}
		\label{Hint}
		\begin{split}
			H_{\rm int}^{(\Lambda)}&:=-g\sum_{x \in \Lambda}\sum_{\mu=1}^3
			\bigl[\Gamma^{(1)}(x)\Gamma^{(1)}(x+e_\mu)+\Gamma^{(2)}(x)\Gamma^{(2)}(x+e_\mu)\bigr] \\
			&\quad -g\sum_{x\in\Lambda}\sum_{\mu=1}^3\sum_{j=1}^3\bigl[S^{(j)}(x)S^{(j)}(x+e_\mu)+S_5^{(j)}(x)S_5^{(j)}(x+e_\mu)\bigr]
		\end{split}
	\end{equation}
	with the coupling constant $g>0$, where
	\begin{equation*}
		\Gamma^{(1)}(x):=\Psi^\dagger(x)\gamma_0\Psi(x),\quad
		\Gamma^{(2)}(x):=\Psi^\dagger(x)i\gamma_0\gamma_5\Psi(x),
	\end{equation*}
	\begin{equation*}
		S^{(j)}(x):=\Psi^\dagger(x)\gamma_0\tau_j\Psi(x)\quad \mbox{and}\quad S_5^{(\ell)}(x):=\Psi^\dagger(x)i\gamma_0\gamma_5\tau_\ell\Psi(x)
	\end{equation*}
	for $j,\ell=1,2,3$, where $\tau_j$ are the Pauli matrices which act only on the flavor indices. The explicit forms are given by
	\begin{equation*}
		\tau_1:=\begin{pmatrix}
			0 & 1 \\ 1 & 0 \end{pmatrix},\quad
		\tau_2:=\begin{pmatrix}
			0 & -i \\ i & 0
		\end{pmatrix}
		\quad \mbox{and} \quad
		\tau_3:=\begin{pmatrix}
			1 & 0 \\ 0 & -1
		\end{pmatrix}.
	\end{equation*}
	In the following, we will use the expressions,
	\begin{equation}
		\label{gamma05}
		\gamma_0=\begin{pmatrix}
			1_2 & 0 \\ 0 & -1_2
		\end{pmatrix}
		\quad \mbox{and} \quad \gamma_5=\begin{pmatrix}
			0 & 1_2 \\ 1_2 & 0 \end{pmatrix},
	\end{equation}
	with the 2-by-2 unit matrix $1_2$. We also impose the anti-periodic boundary condition \cite{GK}
	for the boundary of the kinetic Hamiltonian (\ref{HK}), in order to realize the fermion reflection positivity~\cite{JP}.

	Originally, reflection positivity was introduced in quantum field theory by Osterwalder and Schrader~\cite{OS}, and it is closely related to the existence of a positive self-adjoint transfer matrix~\cite{OsSe}.
	Subsequently, this concept was applied to the study of phase transitions in statistical mechanical models~\cite{FILS,FSS,DLS}.
	In the present paper, reflection positivity is the essential tool to establish the existence of long-range order.

	\section{Continuous chiral symmetry of the Hamiltonian}
	\label{sec:chiralsymm}

	As is well known, the form (\ref{Hint}) of the interaction Hamiltonian $H_{\rm int}^{(\Lambda)}$ is invariant
	under the usual chiral transformation,
	\begin{equation}
		\label{chiraltrans}
		\Psi_{\rm u}(x)\rightarrow e^{i\theta\gamma_5}\Psi_{\rm u}(x), \quad
		\Psi_{\rm d}(x)\rightarrow e^{i\theta\gamma_5}\Psi_{\rm d}(x),
	\end{equation}
	with a real $\theta$ for $x\in\Lambda$. As we will see below, the present Hamiltonian $H^{(\Lambda)}$ of
	(\ref{HamLam}) is also invariant
	under another chiral transformation,
	\begin{equation}
		\label{chiraltrans2}
		\Psi_{\rm u}(x)\rightarrow e^{i\theta\gamma_5}\Psi_{\rm u}(x), \quad
		\Psi_{\rm d}(x)\rightarrow e^{-i\theta\gamma_5}\Psi_{\rm d}(x),
	\end{equation}
	where the two flavors are rotated by the opposite angles. Since $\gamma_0e^{i\theta\gamma_5}\gamma_0=e^{-i\theta\gamma_5}$,
	one notices that the quantity $\Psi^\dagger(x)\gamma_0\tau_1\Psi(y)$ is invariant under this transformation.
	More generally, the operators, $S^{(j)}(x)$ and $S_5^{(j)}(x)$ for $j=1,2$, are all invariant under this transformation.

	The rest of the terms are given by the operators,
	\begin{equation*}
		\Gamma^{(1)}(x)\Gamma^{(1)}(x+e_\mu),\, \Gamma^{(2)}(x)\Gamma^{(2)}(x+e_\mu),
		\, S^{(3)}(x)S^{(3)}(x+e_\mu)\, \mbox{and} \, S_5^{(3)}(x)S_5^{(3)}(x+e_\mu).
	\end{equation*}
	Note that
	\begin{equation*}
		\Gamma^{(1)}(x)=\Psi_{\rm u}^\dagger(x)\gamma_0\Psi_{\rm u}(x)+\Psi_{\rm d}^\dagger(x)\gamma_0\Psi_{\rm d}(x)
	\end{equation*}
	and
	\begin{equation*}
		S_5^{(3)}(x)=\Psi_{\rm u}^\dagger(x)i\gamma_0\gamma_5\Psi_{\rm u}(x)-\Psi_{\rm d}^\dagger(x)i\gamma_0\gamma_5\Psi_{\rm d}(x).
	\end{equation*}
	Under the present chiral transformation (\ref{chiraltrans2}), one has
	\begin{equation*}
		\Gamma^{(1)}(x)\rightarrow \Gamma^{(1)}(x)\cos 2\theta +S_5^{(3)}(x)\sin 2\theta,
	\end{equation*}
	\begin{equation*}
		S_5^{(3)}(x)\rightarrow -\Gamma^{(1)}(x)\sin 2\theta +S_5^{(3)}(x)\cos 2\theta.
	\end{equation*}
	Therefore, the following quantity is invariant under the chiral transformation (\ref{chiraltrans2}):
	\begin{equation*}
		\Gamma^{(1)}(x)\Gamma^{(1)}(y)+S_5^{(3)}(x)S_5^{(3)}(y)
	\end{equation*}
	for $x,y\in\Lambda$. Similarly, the following quantity is also invariant:
	\begin{equation*}
		\Gamma^{(2)}(x)\Gamma^{(2)}(y)+S^{(3)}(x)S^{(3)}(y).
	\end{equation*}
	These observations imply that the interaction Hamiltonian $H_{\rm int}^{(\Lambda)}$ of (\ref{Hint})
	is invariant under the chiral transformation (\ref{chiraltrans2}).
	Since the kinetic Hamiltonian $H_{\rm K}^{(\Lambda)}$ is invariant,
	the present Hamiltonian $H^{(\Lambda)}$ of (\ref{HamLam}) is invariant under the chiral transformation (\ref{chiraltrans2}).

	\section{Spontaneous breaking of both flavor and parity symmetries}
	\label{sec:result}

	The aim of the present paper is to prove the spontaneous breaking of both flavor and parity symmetries.

	We write
	\begin{equation*}
		\omega_0^{(\Lambda)}(\cdots):=\lim_{\beta\nearrow\infty}
		\frac{{\rm Tr}\;(\cdots)\;e^{-\beta H^{(\Lambda)}}}{{\rm Tr}\;e^{-\beta H^{(\Lambda)}}}
	\end{equation*}
	for the expectation value with respect to the ground state without the symmetry breaking field.
	In order to consider a phase transition, we introduce two order parameters
	\begin{equation*}
		O_\Lambda^{(1)}=\frac{1}{|\Lambda|}\sum_{x\in\Lambda}S^{(1)}(x)=\frac{1}{|\Lambda|}\sum_{x\in\Lambda}\Psi^\dagger(x)\gamma_0\tau_1\Psi(x)
	\end{equation*}
	and
	\begin{equation*}
		O_{5,\Lambda}^{(2)}=\frac{1}{|\Lambda|}\sum_{x\in\Lambda}S_5^{(2)}(x)
		=\frac{1}{|\Lambda|}\sum_{x\in\Lambda}\Psi^\dagger(x)i\gamma_0\gamma_5\tau_2\Psi(x).
	\end{equation*}
	By a standard argument (see also Appendix~\ref{ChiralRot}) , it is easy to see that for any $\theta \in\mathbb{R}$,
	\[
	e^{i\theta \gamma_5}\gamma_0e^{-i\theta \gamma_5}
	=\gamma_0\cos(2\theta)-i\gamma_0\gamma_5\sin(2\theta).
	\]
	This relation implies that under  the chiral transformation (\ref{chiraltrans}), $S^{(1)}(x)$ is transformed into $\Psi^\dagger(x)i\gamma_0\gamma_5\tau_1\Psi(x)$.
	Then, the chiral and flavor symmetries of the Hamiltonian $H^{(\Lambda)}$ lead to the following equalities
	\[
	\omega_0^{(\Lambda)}(O_\Lambda^{(1)})=\omega_0^{(\Lambda)}(O_{5,\Lambda}^{(2)})=0,
	\]
	and
	\[
	\omega_0^{(\Lambda)}([O_\Lambda^{(1)}]^2)=\omega_0^{(\Lambda)}([O_{5,\Lambda}^{(2)}]^2).
	\]

	As we will prove below,
	there exists a long-range order in the sense that
	\begin{equation}
		\label{LROO}
		\omega_0^{(\Lambda)}([O_\Lambda^{(1)}]^2)=\omega_0^{(\Lambda)}([O_{5,\Lambda}^{(2)}]^2)>c_0,
	\end{equation}
	where $c_0$ is a strictly positive constant independent of the volume $|\Lambda|$.
	We consider a state \cite{HL,KHL,KT,KT2},
	\begin{equation*}
		\omega^{(\Lambda)}(\cdots):=\mathcal{N}\omega_0^{(\Lambda)}((1+O_\Lambda)(\cdots)(1+O_\Lambda)),
	\end{equation*}
	where $\mathcal{N}$ is the normalization constant, and we have written
	\begin{equation*}
		O_\Lambda:=O_\Lambda^{(1)}+O_{5,\Lambda}^{(2)}.
	\end{equation*}
	This is also a ground-state expectation. To see this, we compute the expectation value of
	$H^{(\Lambda)}-E_0^{(\Lambda)}$, where $E_0^{(\Lambda)}$ is the energy eigenvalue of the ground state.
	We have
	\begin{align*}
		\omega^{(\Lambda)}([H^{(\Lambda)}-E_0^{(\Lambda)}])
		&=\mathcal{N}\omega_0^{(\Lambda)}((1+O_\Lambda)[H^{(\Lambda)}-E_0^{(\Lambda)}](1+O_\Lambda)) \\
		&=\mathcal{N}\omega_0^{(\Lambda)}(O_\Lambda[H^{(\Lambda)}-E_0^{(\Lambda)}]O_\Lambda) \\
		&=\frac{\mathcal{N}}{2}\bigl\{\omega_0^{(\Lambda)}(O_\Lambda[H^{(\Lambda)},O_\Lambda])
		-\omega_0^{(\Lambda)}([H^{(\Lambda)},O_\Lambda]O_\Lambda)\bigr\} \\
		&=\frac{\mathcal{N}}{2}\omega_0^{(\Lambda)}([O_\Lambda, [H^{(\Lambda)},O_\Lambda]])\le \frac{c_1}{|\Lambda|},
	\end{align*}
	where $c_1$ is a positive constant. Thus, the energy is degenerate into the energy of the ground state in the infinite-volume
	limit $|\Lambda|\nearrow \infty$.

	By using the flavor rotational symmetry of the Hamiltonian $H^{(\Lambda)}$ of (\ref{HamLam})
	and the existence of the long-range order (\ref{LROO}), we obtain
	\begin{equation*}
		\omega^{(\Lambda)}(O_\Lambda^{(1)})\ge 2\mathcal{N}c_0\quad \mbox{and}\quad
		\omega^{(\Lambda)}(O_{5,\Lambda}^{(2)})\ge 2\mathcal{N}c_0.
	\end{equation*}
	Thus, the two order parameters show the non-vanishing spontaneous magnetizations in the infinite-volume ground state
	$\omega(\cdots):={\rm weak}^\ast{\rm -}\lim_{\Lambda\nearrow \mathbb{Z}^3}\omega^{(\Lambda)}(\cdots)$.

	These observations imply that the chiral, flavor and parity symmetries are all spontaneously broken in the infinite-volume ground
	state $\omega(\cdots)$.
	\medskip

	Our main result is summarized as follows:

	\begin{theorem}
		\label{Th}
		There exists a positive number $\alpha_0$ small enough such that
		\[
		\lim_{\Lambda \nearrow \mathbb{Z}^3} \omega^{(\Lambda)}(O_\Lambda^{(1)}) >0\quad \mbox{and}\quad
		\lim_{\Lambda \nearrow \mathbb{Z}^3} \omega^{(\Lambda)}(O_{5,\Lambda}^{(2)}) >0
		\]
		for all the hopping parameters $\kappa$ and the coupling constant $g$ of the interaction
		which satisfy $\vert\kappa\vert/g \le \alpha_0$.
	\end{theorem}

	\begin{remark}
		\label{rem.th}
		In passing, we should remark the following:
		\begin{enumerate}
			\item[(i)] In order to prove the violation of the parity,
			we need the two observables because of the symmetries of the present system.
			Actually, as shown in Appendix~\ref{ChiralRot}, a chiral rotation yields the transformation
			$\Psi^\dagger(x)i\gamma_0\gamma_5\tau_2\Psi(x) \rightarrow \Psi^\dagger(x)\gamma_0\tau_2\Psi(x)$.

			\item[(ii)]
			{It may seem that another transformation could express both order parameters as scalar bilinears.
				For instance, under the chiral-flavor rotation $\Psi=U\Psi'$, where $U=e^{-i\pi\gamma_5 \tau_2/4}$, one has
				\[
				\Psi^\dagger \gamma_0\tau_1\Psi=(\Psi')^\dagger\gamma_0\tau_1\Psi',
				\qquad
				\Psi^\dagger \gamma_0i\gamma_5\tau_2\Psi=(\Psi')^\dagger\gamma_0\Psi'.
				\]
				However, the original parity transformation acts in the new variables as
				\[
				\Psi'(x)\mapsto
				U^\dagger\gamma_0U\,\Psi'(-x)
				=-i\gamma_0\gamma_5\tau_2\Psi'(-x).
				\]
				Thus $(\Psi')^\dagger\gamma_0\Psi'$ remains odd under this parity.
				Theorem~\ref{Th} establishes the breaking of this fixed parity, and it does not assert the breaking of every chiral conjugate parity.
			}

			\item[(iii)] We have not been able to prove that the state $\omega(\cdots)$ is pure\footnote{In general, it is quite difficult to prove whether a given state is pure or not.
				Here, the term ``pure state'' refers to the concept in the context of infinite systems~\cite{BR1,BR2}, rather than its usage in usual quantum mechanics with finitely many degrees of freedom.}.
			Therefore, when we consider the pure state decomposition for the state $\omega(\cdots)$,
			it is not obvious that there exists a pure state which shows a non-vanishing spontaneous magnetization for both of the order
			parameters. Physically, since each of the two flavors can take any direction for their chiral rotations,
			one can expect the existence of a pure state having the desired symmetry breaking character.
			In other words, the present internal degrees of freedom allow to yield the two non-vanishing spontaneous magnetizations.
			In fact, these magnetizations can be expected from the commutativity $[\gamma_0\tau_1,i\gamma_0\gamma_5\tau_2]=0$.
			Namely, the first spontaneous magnetization does not seem to prevent the second observable from magnetizing.
			Thus, the assumption of the two flavors is essential for our proof of the violation of the parity.
		\end{enumerate}
	\end{remark}
	\section{Reflection positivity: Preliminary}
	\label{sec:Reflec.Preli}

	Let $\Lambda' \subset \Lambda$ be a subset and $\mathcal{A}(\Lambda')$ be the algebra generated by
	$\psi_{\rm f}^{(i)}(x)$ and $[\psi_{\rm f'}^{(j)}(y)]^\dagger$ for  $x, y \in \Lambda'$, ${\rm f,f'}\in\{{\rm u,d}\}$
	and $i,j\in\{1,2,3,4\}$.
	Since our $\Lambda$ is symmetric with respect to a plane with the periodic boundary condition,
	there are a natural decomposition $\Lambda = \Lambda_- \cup \Lambda_+$ with $\Lambda_- \cap \Lambda_+ = \emptyset$
	and a reflection map $r: \Lambda_{\pm} \to \Lambda_{\mp}$ satisfying $r(\Lambda_\pm) = \Lambda_\mp$.
	We write $\mathcal{A} = \mathcal{A}(\Lambda)$ and $\mathcal{A}_\pm = \mathcal{A}(\Lambda_\pm)$.
	The reflection has an anti-linear representation $\vartheta: \mathcal{A}_\pm \to \mathcal{A}_\mp$ requiring \cite{JP}
	\begin{align*}
		&\vartheta(\psi_{\rm f}^{(i)}(x)) = \psi_{\rm f}^{(i)}(\vartheta(x)), \quad
		\vartheta([\psi_{\rm f}^{(i)}(x)]^\dagger)= [\psi_{\rm f}^{(i)}(\vartheta(x))]^\dagger,\\
		&\vartheta(AB) = \vartheta(A)\vartheta(B), \quad \vartheta(A)^\dagger = \vartheta(A^\dagger) \quad \text{for } A, B \in \mathcal{A}.
	\end{align*}

	For $x \in \Lambda $, we introduce Majorana fermion operators $\xi_{\rm f}^{(i)}(x), \eta_{\rm f}^{(i)}(x)$ by
	\begin{equation}
		\label{Majoranapsi}
		\xi_{\rm f}^{(i)}(x) := [\psi_{\rm f}^{(i)}(x)]^\dagger + \psi_{\rm f}^{(i)}(x), \quad
		\eta_{\rm f}^{(i)}(x) := i\{[\psi_{\rm f}^{(i)}(x)]^\dagger - \psi_{\rm f}^{(i)}(x)\},
	\end{equation}
	or equivalently,
	\begin{equation}
		\label{psiMajorana}
		\psi_{\rm f}^{(i)}(x) = \frac{1}{2}[\xi_{\rm f}^{(i)}(x) + i\eta_{\rm f}^{(i)}(x)], \quad
		[\psi_{\rm f}^{(i)}(x)]^\dagger = \frac{1}{2}[\xi_{\rm f}^{(i)}(x) - i\eta_{\rm f}^{(i)}(x)].
	\end{equation}
	These satisfy $[\xi_{\rm f}^{(i)}(x)]^\dagger =\xi_{\rm f}^{(i)}(x)$,
	$[\eta_{\rm f}^{(i)}(x)]^\dagger = \eta_{\rm f}^{(i)}(x)$, and the anti-commutation relations
	\begin{align*}
		\{\xi_{\rm f}^{(i)}(x), \xi_{\rm f'}^{(j)}(y)\} &= 2\delta_{x, y}\delta_{i,j}\delta_{\rm f,f'},  \quad
		\{{\eta}_{\rm f}^{(i)}(x), {\eta}_{\rm f'}^{(j)}(y)\} =2\delta_{x, y}\delta_{i,j}\delta_{\rm f,f'}, \\
		\{\xi_{\rm f}^{(i)}(x), \eta_{\rm f'}^{(j)}(y)\} &=0.
	\end{align*}

	Next, following the idea of \cite{FSS}, we will introduce certain functions, $h_\mu$, on the lattice $\Lambda$, and
	rewrite the interaction Hamiltonian. For this purpose, we first decompose the interaction Hamiltonian into two parts,
	\begin{equation*}
		H_{\rm int}^{(\Lambda)}=H_{\rm int,R}^{(\Lambda)}+H_{\rm int, I}^{(\Lambda)},
	\end{equation*}
	where
	\begin{equation}
		\label{HintR}
		\begin{split}
			H_{\rm int, R}^{(\Lambda)}:=-g\sum_{x \in \Lambda}\sum_{\mu=1}^3
			&\bigl[\Gamma^{(1)}(x)\Gamma^{(1)}(x+e_\mu)+S^{(1)}(x)S^{(1)}(x+e_\mu) \\
			&\quad  +S^{(3)}(x)S^{(3)}(x+e_\mu)+S_5^{(2)}(x)S_5^{(2)}(x+e_\mu)\bigr]
		\end{split}
	\end{equation}
	and
	\begin{equation}
		\label{HintI}
		\begin{split}
			H_{\rm int, I}^{(\Lambda)}:=-g\sum_{x\in\Lambda}\sum_{\mu=1}^3
			[&\Gamma^{(2)}(x)\Gamma^{(2)}(x+e_\mu)+S^{(2)}(x)S^{(2)}(x+e_\mu)\\
			&  +S_5^{(1)}(x)S_5^{(1)}(x+e_\mu)+S_5^{(3)}(x)S_5^{(3)}(x+e_\mu)\bigr].
		\end{split}
	\end{equation}
	Note that
	\begin{equation*}
		\sum_{x \in \Lambda}\sum_{\mu=1}^3\Gamma^{(1)}(x)\Gamma^{(1)}(x+e_\mu)
		=-\frac{1}{2}\sum_{x\in\Lambda}\sum_{\mu=1}^3 [\Gamma^{(1)}(x)-\Gamma^{(1)}(x+e_\mu)]^2
		+3\sum_{x\in\Lambda}[\Gamma^{(1)}(x)]^2.
	\end{equation*}
	Therefore, one has
	\begin{equation}
		\label{HintR2}
		\begin{split}
			H_{\rm int, R}^{(\Lambda)}&=\frac{g}{2}\sum_{x \in \Lambda}\sum_{\mu=1}^3
			\bigl\{\bigl[\Gamma^{(1)}(x)-\Gamma^{(1)}(x+e_\mu)\bigr]^2+\bigl[S^{(1)}(x)-S^{(1)}(x+e_\mu)\bigr]^2\\
			&  \qquad +\bigl[S^{(3)}(x)-S^{(3)}(x+e_\mu)\bigr]^2+\bigl[S_5^{(2)}(x)-S_5^{(2)}(x+e_\mu)\bigr]^2\bigr\} \\
			& \qquad -3g\sum_{x\in\Lambda}\bigl\{[\Gamma^{(1)}(x)]^2+[S^{(1)}(x)]^2+[S^{(3)}(x)]^2+[S_5^{(2)}(x)]^2\bigr\}.
		\end{split}
	\end{equation}
	Similarly, one has
	\begin{equation*}
		\sum_{x \in \Lambda}\sum_{\mu=1}^3\Gamma^{(2)}(x)\Gamma^{(2)}(x+e_\mu)=
		\frac{1}{2}\sum_{x\in\Lambda}\sum_{\mu=1}^3 [\Gamma^{(2)}(x)+\Gamma^{(2)}(x+e_\mu)]^2
		-3\sum_{x\in\Lambda}[\Gamma^{(2)}(x)]^2.
	\end{equation*}
	Therefore,
	\begin{equation}
		\label{HintI2}
		\begin{split}
			H_{\rm int, I}^{(\Lambda)}=&-\frac{g}{2}\sum_{x\in\Lambda}\sum_{\mu=1}^3\bigl\{
			[\Gamma^{(2)}(x)+\Gamma^{(2)}(x+e_\mu)]^2+[S^{(2)}(x)+S^{(2)}(x+e_\mu)]^2\\
			& +[S_5^{(1)}(x)+S_5^{(1)}(x+e_\mu)]^2+[S_5^{(3)}(x)+S_5^{(3)}(x+e_\mu)]^2\bigr\}\\
			&+3g\sum_{x\in\Lambda}\bigl\{[\Gamma^{(2)}(x)]^2+[S^{(2)}(x)]^2+[S_5^{(1)}(x)]^2+[S_5^{(3)}(x)]^2\bigr\}
		\end{split}
	\end{equation}

	Let $h_\mu(x)$ be a real-valued function on the lattice $\Lambda$ for $\mu=1,2,3$, and we introduce \cite{DLS,FSS}
	\begin{equation}
		\label{HintR2hf}
		\begin{split}
			H_{\rm int, R}^{(\Lambda)}(h):=&\frac{g}{2}\sum_{x \in \Lambda}\sum_{\mu=1}^3
			\bigl\{\bigl[\Gamma^{(1)}(x)-\Gamma^{(1)}(x+e_\mu)+h_\mu(x)\bigr]^2
			+\bigl[S^{(1)}(x)-S^{(1)}(x+e_\mu)\bigr]^2\\
			&  +\bigl[S^{(3)}(x)-S^{(3)}(x+e_\mu)\bigr]^2+\bigl[S_5^{(2)}(x)-S_5^{(2)}(x+e_\mu)\bigr]^2\bigr\}\\
			&  -3g\sum_{x\in\Lambda}\bigl\{[\Gamma^{(1)}(x)]^2+[S^{(1)}(x)]^2+[S^{(3)}(x)]^2+[S_5^{(2)}(x)]^2\bigr\}.
		\end{split}
	\end{equation}
	We write
	\begin{equation}
		\label{Hinthf}
		H_{\rm int}^{(\Lambda)}(h):=H_{\rm int,R}^{(\Lambda)}(h)+H_{\rm int,I}^{(\Lambda)}.
	\end{equation}
	Clearly, when $h_\mu=0$ for all $\mu=1,2,3$, this equals the interaction Hamiltonian $H_{\rm int}^{(\Lambda)}$.
	We also write
	\begin{equation}
		\label{Hmhf}
		H^{(\Lambda)}(h):=H_{\rm K}^{(\Lambda)}+H_{\rm int}^{(\Lambda)}(h)
	\end{equation}
	for the whole Hamiltonian.

	\section{Reflection with respect to the plane $x^{(1)}=1/2$}
	\label{sec:x1=1/2}

	Let us consider first the reflection with respect to the $x^{(1)}=1/2$ plane.
	By this plane, we divide our finite lattice $\Lambda$ into two parts,
	\[
	\Lambda_- :=\{x \in \Lambda \colon -L+1 \le x^{(1)} \le 0\}
	\quad \mbox{and}\quad
	\Lambda_+ := \{x  \in \Lambda \colon 1 \le x^{(1)} \le L\}.
	\]
	In order to show that the present Hamiltonian has a reflection positivity with respect to this plane, we need some preparations.

	We first introduce a transformation,
	\begin{equation*}
		\Psi_{\rm u}(x)\rightarrow \gamma_0\Psi_{\rm u}(x), \quad \Psi_{\rm d}(x)\rightarrow \Psi_{\rm d}(x),
	\end{equation*}
	and write $U(\gamma_0)$ for the unitary transformation on the fermion Fock space.
	We also introduce a transformation,
	\begin{equation*}
		\Psi(x)\rightarrow e^{i\pi x^{(2)}/2}\Psi(x),
	\end{equation*}
	and write $U_2$ for the corresponding unitary transformation on the fermion Fock space.

	Let us consider first the kinetic Hamiltonian $H_{\rm K}^{(\Lambda)}$ of (\ref{HK}) about these two transformations.
	We have written $H_{\rm K, f}^{(\Lambda)}$ for the kinetic Hamiltonian $H_{\rm K}^{(\Lambda)}$ (\ref{HK}) for
	the flavor ${\rm f}\in \{{\rm u,d}\}$. Clearly, the first transformation $U(\gamma_0)$
	changes the sign of the kinetic Hamiltonian for the flavor u, i.e.,
	\begin{equation*}
		U^\dagger(\gamma_0)H_{\rm K, u}^{(\Lambda)}U(\gamma_0)=-H_{\rm K, u}^{(\Lambda)}.
	\end{equation*}
	Therefore, by these two transformations, $U(\gamma_0)$ and $U_2$,
	the kinetic Hamiltonians $H_{\rm K,f}^{(\Lambda)}$ for the flavors ${\rm f}={\rm u, d}$ are transformed into the following forms:
	\begin{equation}
		\label{tildeHKu}
		\begin{split}
			\tilde{H}_{\rm K, u}^{(\Lambda)}:=&[U(\gamma_0)U_2]^\dagger H_{\rm K, u}^{(\Lambda)}U(\gamma_0)U_2 \\
			=&-i\kappa \sum_{x\in\Lambda\subset\mathbb{Z}^3}
			\Bigl\{[\Psi_{\rm u}^\dagger(x)\alpha_1\Psi_{\rm u}(x+e_1)-\Psi_{\rm u}^\dagger(x+e_1)\alpha_1\Psi_{\rm u}(x)] \\
			&\qquad \qquad +i[\Psi_{\rm u}^\dagger(x)\alpha_2\Psi_{\rm u}(x+e_2)+\Psi_{\rm u}^\dagger(x+e_2)\alpha_2\Psi_{\rm u}(x)] \\
			&\qquad \qquad +[\Psi_{\rm u}^\dagger(x)\alpha_3\Psi_{\rm u}(x+e_3)-\Psi_{\rm u}^\dagger(x+e_3)\alpha_3\Psi_{\rm u}(x)]\Bigr\}
		\end{split}
	\end{equation}
	and
	\begin{equation}
		\label{tildeHKd}
		\begin{split}
			\tilde{H}_{\rm K,d}^{(\Lambda)}&:=[U(\gamma_0)U_2]^\dagger H_{\rm K,d}^{(\Lambda)}U(\gamma_0)U_2\\
			&=i\kappa \sum_{x\in\Lambda\subset\mathbb{Z}^3}
			\Bigl\{[\Psi_{\rm d}^\dagger(x)\alpha_1\Psi_{\rm d}(x+e_1)-\Psi_{\rm d}^\dagger(x+e_1)\alpha_1\Psi_{\rm d}(x)]\\
			& \qquad \qquad +i[\Psi_{\rm d}^\dagger(x)\alpha_2\Psi_{\rm d}(x+e_2)+\Psi_{\rm d}^\dagger(x+e_2)\alpha_2\Psi_{\rm d}(x)]\\
			& \qquad \qquad +[\Psi_{\rm d}^\dagger(x)\alpha_3\Psi_{\rm d}(x+e_3)-\Psi_{\rm d}^\dagger(x+e_3)\alpha_3\Psi_{\rm d}(x)]\Bigr\}.
		\end{split}
	\end{equation}
	When we use the real representation for the Dirac fermion field $\Psi(x)$, these right-hand sides become
	pure imaginary hermitian by the expressions of the matrices, $\alpha_1, \alpha_2, \alpha_3$. This property
	is crucial for the reflection positivity \cite{JP}.

	In order to deal with the interaction Hamiltonian, we note that
	\begin{equation*}
		\tilde{\Gamma}^{(1)}(x):=[U(\gamma_0)U_2]^\dagger\Gamma^{(1)}(x)U(\gamma_0)U_2=\Psi^\dagger(x)\gamma_0\Psi(x)=\Gamma^{(1)}(x),
	\end{equation*}
	\begin{equation}
		\label{tildeGamma2}
		\tilde{\Gamma}^{(2)}(x):=[U(\gamma_0)U_2]^\dagger\Gamma^{(2)}(x)U(\gamma_0)U_2=-\Psi^\dagger(x)i\gamma_0\gamma_5\tau_3\Psi(x)
		=-S_5^{(3)}(x),
	\end{equation}
	\begin{equation*}
		\tilde{S}^{(1)}(x):=[U(\gamma_0)U_2]^\dagger S^{(1)}(x)U(\gamma_0)U_2=\Psi^\dagger(x)\tau_1\Psi(x),
	\end{equation*}
	\begin{equation*}
		\tilde{S}^{(2)}(x):=[U(\gamma_0)U_2]^\dagger S^{(2)}(x)U(\gamma_0)U_2=\Psi^\dagger(x)\tau_2\Psi(x),
	\end{equation*}
	\begin{equation*}
		\tilde{S}^{(3)}(x):=[U(\gamma_0)U_2]^\dagger S^{(3)}(x)U(\gamma_0)U_2=\Psi^\dagger(x)\gamma_0\tau_3\Psi(x)=S^{(3)}(x),
	\end{equation*}
	and
	\begin{equation*}
		\tilde{S}_5^{(1)}(x):=[U(\gamma_0)U_2]^\dagger S_5^{(1)}(x)U(\gamma_0)U_2=-\Psi^\dagger(x)\gamma_5\tau_2\Psi(x),
	\end{equation*}
	\begin{equation*}
		\tilde{S}_5^{(2)}(x):=[U(\gamma_0)U_2]^\dagger S_5^{(2)}(x)U(\gamma_0)U_2=\Psi^\dagger(x)\gamma_5\tau_1\Psi(x)
	\end{equation*}
	and
	\begin{equation}
		\label{tildeS53}
		\tilde{S}_5^{(3)}(x):=[U(\gamma_0)U_2]^\dagger S_5^{(3)}(x)U(\gamma_0)U_2=-\Psi^\dagger(x)i\gamma_0\gamma_5\Psi(x)
		=-\Gamma^{(2)}(x).
	\end{equation}
	Then
	\begin{equation*}
		\tilde{H}_{\rm int}^{(\Lambda)}(h):=[U(\gamma_0)U_2]^\dagger H_{\rm int}^{(\Lambda)}(h)U(\gamma_0)U_2
		=\tilde{H}_{\rm int,R}^{(\Lambda)}(h)+\tilde{H}_{{\rm int},{\rm I}}^{(\Lambda)},
	\end{equation*}
	where we have written
	\begin{equation}
		\label{tildeHintRhf}
		\begin{split}
			\tilde{H}_{\rm int,R}^{(\Lambda)}(h):=&[U(\gamma_0)U_2]^\dagger H_{\rm int,R}^{(\Lambda)}(h)U(\gamma_0)U_2 \\
			&=\frac{g}{2}\sum_{x \in \Lambda}\sum_{\mu=1}^3
			\bigl\{\bigl[\tilde{\Gamma}^{(1)}(x)-\tilde{\Gamma}^{(1)}(x+e_\mu)+h_\mu(x)\bigr]^2
			+\bigl[\tilde{S}^{(1)}(x)-\tilde{S}^{(1)}(x+e_\mu)\bigr]^2 \\
			& \quad +\bigl[\tilde{S}^{(3)}(x)-\tilde{S}^{(3)}(x+e_\mu)\bigr]^2
			+\bigl[\tilde{S}_5^{(2)}(x)-\tilde{S}_5^{(2)}(x+e_\mu)\bigr]^2\bigr\} \\
			& \quad -3g\sum_{x\in\Lambda}\bigl\{[\tilde{\Gamma}^{(1)}(x)]^2+[\tilde{S}^{(1)}(x)]^2+[\tilde{S}^{(3)}(x)]^2
			+[\tilde{S}_5^{(2)}(x)]^2\bigr\}
		\end{split}
	\end{equation}
	and
	\begin{equation}
		\label{tildeHintI}
		\begin{split}
			\tilde{H}_{{\rm int},{\rm I}}^{(\Lambda)}&:=[U(\gamma_0)U_2]^\dagger H_{{\rm int},{\rm I}}^{(\Lambda)}U(\gamma_0)U_2 \\
			&=-\frac{g}{2}\sum_{x\in\Lambda}\sum_{\mu=1}^3\bigl\{
			[\tilde{\Gamma}^{(2)}(x)+\tilde{\Gamma}^{(2)}(x+e_\mu)]^2+[\tilde{S}^{(2)}(x)+\tilde{S}^{(2)}(x+e_\mu)]^2 \\
			& \qquad\qquad +[\tilde{S}_5^{(1)}(x)+\tilde{S}_5^{(1)}(x+e_\mu)]^2+[\tilde{S}_5^{(3)}(x)+\tilde{S}_5^{(3)}(x+e_\mu)]^2\bigr\} \\
			&+3g\sum_{x\in\Lambda}\bigl\{[\tilde{\Gamma}^{(2)}(x)]^2+[\tilde{S}^{(2)}(x)]^2+[\tilde{S}_5^{(1)}(x)]^2+[\tilde{S}_5^{(3)}(x)]^2\bigr\}.
		\end{split}
	\end{equation}
	The whole Hamiltonian $H^{(\Lambda)}(h)$ of (\ref{Hmhf}) is transformed into
	\begin{equation}
		\label{tildeHmhf}
		\begin{split}
			\tilde{H}^{(\Lambda)}(h)&:=[U(\gamma_0)U_2]^\dagger H^{(\Lambda)}(h)U(\gamma_0)U_2 \\
			&=\tilde{H}_{\rm K,u}^{(\Lambda)}+\tilde{H}_{\rm K,d}^{(\Lambda)}
			+\tilde{H}_{\rm int,R}^{(\Lambda)}(h)+\tilde{H}_{\rm int, I}^{(\Lambda)}.
		\end{split}
	\end{equation}

	\subsection{Two unitary transformations}

	Further, we will use two unitary transformations below.
	First we define $U(\alpha_1)$ as follows:
	\begin{equation}
		\label{Ualpha1}
		[U(\alpha_1)]^\dagger \Psi_{\rm u}(x)U(\alpha_1)=\begin{cases}
			-\alpha_1\Psi_{\rm u}(x) & \mbox{for \ } x\in\Lambda_+;\\
			\Psi_{\rm u}(x) & \mbox{for \ } x\in\Lambda_-,
		\end{cases}
	\end{equation}
	and
	\begin{equation}
		\label{Ualpha1d}
		[U(\alpha_1)]^\dagger \Psi_{\rm d}(x)U(\alpha_1)=\begin{cases}
			\Psi_{\rm d}(x) & \mbox{for \ } x\in\Lambda_+;\\
			\alpha_1\Psi_{\rm d}(x) & \mbox{for \ } x\in\Lambda_-.
		\end{cases}
	\end{equation}

	We also introduce \cite{FILS,GK}
	\begin{equation}
		\label{uxsigma}
		u_{\rm PH,f}^{(i)}(x):=\left[\prod_{\substack{y\in\Lambda,\; j\in\{1,2,3,4\},\; {\rm f'}\in\{{\rm u,d}\}
				: \\ (y,j,{\rm f'})\ne (x,i,{\rm f})\; }}
		(-1)^{n_{\rm f'}^{(j)}(y)}\right]\bigl\{[\psi_{\rm f}^{(i)}(x)]^\dagger+\psi_{\rm f}^{(i)}(x)\bigr\},
	\end{equation}
	where we have written
	\begin{equation*}
		n_{\rm f'}^{(j)}(y):=[\psi_{\rm f'}^{(j)}(y)]^\dagger\psi_{\rm f'}^{(j)}(y)
	\end{equation*}
	for $y\in\Lambda$ and ${\rm f'}\in\{{\rm u,d}\}$. Then, one has
	\begin{equation*}
		[u_{\rm PH,f}^{(i)}(x)]^\dagger \psi_{\rm f'}^{(j)}(y)u_{\rm PH,f}^{(i)}(x)=
		\begin{cases}
			[\psi_{\rm f}^{(i)}(x)]^\dagger, & \mbox{for}\; (y,j,{\rm f'})=(x,i,{\rm f}) ;\\
			\psi_{\rm f'}^{(j)}(y), & \mbox{otherwise}.
		\end{cases}
	\end{equation*}
	By using these operators, we define a particle-hole transformation on a sublattice by \cite{FILS,GK}
	\begin{equation*}
		U_{\rm odd,u}:=\prod_{x\in\Lambda_{\rm odd}}\prod_{j\in\{1,2,3,4\}}u_{\rm PH,u}^{(j)}(x),
	\end{equation*}
	where we have written
	\begin{equation*}
		\Lambda_{\rm odd}:=\{x\in\Lambda\; |\; x^{(1)}+x^{(2)}+x^{(3)}={\rm odd}\}.
	\end{equation*}
	For all $i\in\{1,2,3,4\}$, one has
	\begin{equation}
		\label{UPH}
		(U_{\rm odd,u})^\dagger \psi_{\rm u}^{(i)}(x)U_{\rm odd,u}=
		\begin{cases}
			[\psi_{\rm u}^{(i)}(x)]^\dagger & \mbox{for \ }x\in\Lambda_{\rm odd},\\
			\psi_{\rm u}^{(i)}(x) & \mbox{for \ }x\in\Lambda\backslash\Lambda_{\rm odd}.
		\end{cases}
	\end{equation}
	Similarly, we define
	\begin{equation*}
		U_{\rm even,d}:=\prod_{x\in\Lambda_{\rm even}}\prod_{j\in\{1,2,3,4\}}u_{\rm PH,d}^{(j)}(x).
	\end{equation*}
	for the flavor d and the sublattice $\Lambda_{\rm even}:=\Lambda\backslash\Lambda_{\rm odd}$.
	Further, we write
	\begin{equation*}
		U_{\rm stagg}:=U_{\rm odd,u}U_{\rm even,d},
	\end{equation*}
	\begin{equation*}
		\tilde{U}_1:=U(\alpha_1)U_{\rm stagg},
	\end{equation*}
	and
	\begin{equation*}
		\hat{H}^{(\Lambda)}(h):=(\tilde{U}_1)^\dagger\tilde{H}^{(\Lambda)}(h)\tilde{U}_1
	\end{equation*}
	for the transformed Hamiltonian from (\ref{tildeHmhf}).
	We want to decompose this Hamiltonian into some parts \cite{GK} in the following way.

	\subsection{Kinetic Hamiltonian}

	Let us consider the kinetic Hamiltonians  (\ref{tildeHKu}) and (\ref{tildeHKd}).
	We first deal with $\tilde{H}_{\rm K,u}^{(\Lambda)}$ of (\ref{tildeHKu}). It can be written
	\begin{equation*}
		\tilde{H}_{\rm K,u}^{(\Lambda)}=\sum_{\mu=1}^3 \tilde{H}_{{\rm K,u},\mu}^{(\Lambda)}
	\end{equation*}
	with
	\begin{equation*}
		\tilde{H}_{{\rm K,u},\mu}^{(\Lambda)}:=-i\kappa\sum_{x\in\Lambda}[\Psi_{\rm u}^\dagger(x)\alpha_\mu\Psi_{\rm u}(x+e_\mu)
		-\Psi_{\rm u}^\dagger(x+e_\mu)\alpha_\mu\Psi_{\rm u}(x)]\quad \mbox{for \ } \mu=1,3,
	\end{equation*}
	and
	\begin{equation*}
		\tilde{H}_{{\rm K,u},2}^{(\Lambda)}:=\kappa\sum_{x\in\Lambda}[\Psi_{\rm u}^\dagger(x)\alpha_2\Psi_{\rm u}(x+e_2)
		+\Psi_{\rm u}^\dagger(x+e_2)\alpha_2\Psi_{\rm u}(x)].
	\end{equation*}
	Clearly, the hopping Hamiltonians in the second and third directions can be decomposed into two parts as follows:
	\begin{equation*}
		\tilde{H}_{{\rm K,u},2}^{(\Lambda)}=\tilde{H}_{{\rm K,u},2}^++\tilde{H}_{{\rm K,u},2}^-
	\end{equation*}
	with
	\begin{equation}
		\label{tildeHK2pm}
		\tilde{H}_{{\rm K,u},2}^\pm:=\kappa\sum_{x\in\Lambda_\pm}[\Psi_{\rm u}^\dagger(x)\alpha_2\Psi_{\rm u}(x+e_2)
		+\Psi_{\rm u}^\dagger(x+e_2)\alpha_2\Psi_{\rm u}(x)]
	\end{equation}
	and
	\begin{equation*}
		\tilde{H}_{{\rm K,u},3}^{(\Lambda)}=\tilde{H}_{{\rm K,u},3}^++\tilde{H}_{{\rm K,u},3}^-
	\end{equation*}
	with
	\begin{equation}
		\label{tildeHK3pm}
		\tilde{H}_{{\rm K,u},3}^\pm:=-i\kappa\sum_{x\in\Lambda_\pm}[\Psi_{\rm u}^\dagger(x)\alpha_3\Psi_{\rm u}(x+e_3)
		-\Psi_{\rm u}^\dagger(x+e_3)\alpha_3\Psi_{\rm u}(x)].
	\end{equation}
	The kinetic term in the first direction is decomposed into three parts as follows:
	\begin{equation*}
		\tilde{H}_{{\rm K,u},1}^{(\Lambda)}=\tilde{H}_{{\rm K,u},1}^++\tilde{H}_{{\rm K,u},1}^-+\tilde{H}_{{\rm K,u},1}^0,
	\end{equation*}
	where
	\begin{equation}
		\label{tildeHK1+}
		\tilde{H}_{{\rm K,u},1}^+:=-i\kappa\sum_{x\in\Lambda_+\; :\; x^{(1)}\ne L}
		[\Psi_{\rm u}^\dagger(x)\alpha_1\Psi_{\rm u}(x+e_1)-\Psi^\dagger_{\rm u}(x+e_1)\alpha_1\Psi_{\rm u}(x)],
	\end{equation}
	\begin{equation}
		\label{tildeHK1-}
		\tilde{H}_{{\rm K,u},1}^-:=-i\kappa\sum_{x\in\Lambda_-\; :\; x^{(1)}\ne 0}
		[\Psi_{\rm u}^\dagger(x)\alpha_1\Psi_{\rm u}(x+e_1)-\Psi_{\rm u}^\dagger(x+e_1)\alpha_1\Psi_{\rm u}(x)],
	\end{equation}
	and
	\begin{equation}
		\label{tildeHK10}
		\begin{split}
			\tilde{H}_{{\rm K,u},1}^0:=&
			-i\kappa\sum_{x\in\Lambda\; :\; x^{(1)}=0}[\Psi_{\rm u}^\dagger(x)\alpha_1\Psi_{\rm u}(x+e_1)
			-\Psi_{\rm u}^\dagger(x+e_1)\alpha_1\Psi_{\rm u}(x)] \\
			&-i\kappa\sum_{x\in\Lambda\; :\: x^{(1)}=L}[\Psi_{\rm u}^\dagger(x_L^-)\alpha_1\Psi_{\rm u}(x)
			-\Psi_{\rm u}^\dagger(x)\alpha_1\Psi_{\rm u}(x_L^-)],
		\end{split}
	\end{equation}
	where $x_L^-:=(-L+1,x^{(2)},x^{(3)})$, and we have used the anti-periodic boundary condition for the fermions.


	By using the unitary transformation $U(\alpha_1)$ of (\ref{Ualpha1}), we have
	\begin{align*}
		[U(\alpha_1)]^\dagger\tilde{H}_{{\rm K,u},1}^0U(\alpha_1)=&
		i\kappa\sum_{x\in\Lambda\; :\; x^{(1)}=0}[\Psi_{\rm u}^\dagger(x)\Psi_{\rm u}(x+e_1)-\Psi_{\rm u}^\dagger(x+e_1)\Psi_{\rm u}(x)] \\
		&+i\kappa\sum_{x\in\Lambda\; :\: x^{(1)}=L}[\Psi_{\rm u}^\dagger(x_L^-)\Psi_{\rm u}(x)-\Psi_{\rm u}^\dagger(x)\Psi_{\rm u}(x_L^-)],
	\end{align*}
	where we have used $\alpha_1$ is self-adjoint, and $(\alpha_1)^2=1$.
	In passing, we note the following:
	The negative sign in front of the right-hand side of (\ref{tildeHKu}) is canceled out with
	the negative sign in the right-hand side of (\ref{Ualpha1}). On the other hand, in the case of the flavor d of (\ref{tildeHKd}),
	both of the corresponding two signs are plus. Therefore, we obtain the same result in the case of the flavor d.
	Namely, for the flavor ${\rm d}$, one has
	\begin{align*}
		[U(\alpha_1)]^\dagger\tilde{H}_{{\rm K,d},1}^0U(\alpha_1)=&
		i\kappa\sum_{x\in\Lambda\; :\; x^{(1)}=0}[\Psi_{\rm d}^\dagger(x)\Psi_{\rm d}(x+e_1)-\Psi_{\rm d}^\dagger(x+e_1)\Psi_{\rm d}(x)] \\
		&+i\kappa\sum_{x\in\Lambda\; :\: x^{(1)}=L}[\Psi_{\rm d}^\dagger(x_L^-)\Psi_{\rm d}(x)-\Psi_{\rm d}^\dagger(x)\Psi_{\rm d}(x_L^-)]
	\end{align*}
	{from} (\ref{tildeHKd}) and (\ref{Ualpha1d}).

	By using the Majorana fermions of (\ref{Majoranapsi}), one has
	\begin{equation*}
		[\psi_{\rm u}^{(i)}(x)]^\dagger \psi_{\rm u}^{(i)}(y)-[\psi_{\rm u}^{(i)}(y)]^\dagger\psi_{\rm u}^{(i)}(x)
		=\frac{1}{2}[\xi_{\rm u}^{(i)}(x)\xi_{\rm u}^{(i)}(y)+\eta_{\rm u}^{(i)}(x)\eta_{\rm u}^{(i)}(y)]\quad \mbox{for \ } x\ne y.
	\end{equation*}
	Therefore, we have
	\begin{equation}
		\label{Ualpha1tildeHK1}
		\begin{split}
			[U(\alpha_1)]^\dagger\tilde{H}_{{\rm K,u},1}^0U(\alpha_1)=&
			\frac{i\kappa}{2}\sum_{x\in\Lambda\; :\; x^{(1)}=0}\sum_{i=1}^4[\xi_{\rm u}^{(i)}(x)\xi_{\rm u}^{(i)}(x+e_1)
			+\eta_{\rm u}^{(i)}(x)\eta_{\rm u}^{(i)}(x+e_1)] \\
			&+\frac{i\kappa}{2}\sum_{x\in\Lambda\; :\: x^{(1)}=L}\sum_{i=1}^4
			[\xi_{\rm u}^{(i)}(x_L^-)\xi_{\rm u}^{(i)}(x)+\eta_{\rm u}^{(i)}(x_L^-)\eta_{\rm u}^{(i)}(x)].
		\end{split}
	\end{equation}
	From the representations (\ref{Majoranapsi}) of the Majorana fermions and (\ref{UPH}), one has
	\begin{equation*}
		(U_{\rm odd,u})^\dagger \xi_{\rm u}^{(i)}(x) U_{\rm odd,u}=\xi_{\rm u}^{(i)}(x)\quad \mbox{for all \ } x\in\Lambda \ \mbox{and \ } i=1,2,3,4,
	\end{equation*}
	and
	\begin{equation*}
		(U_{\rm odd,u})^\dagger \eta_{\rm u}^{(i)}(x)U_{\rm odd,u}=
		\begin{cases}
			-\eta_{\rm u}^{(i)}(x) & \mbox{for \ } x\in\Lambda_{\rm odd};\\
			\eta_{\rm u}^{(i)}(x) & \mbox{for \ } x\in\Lambda\backslash\Lambda_{\rm odd}
		\end{cases}
	\end{equation*}
	for all $i=1,2,3,4$. By combining this with $\tilde{U}_1=U(\alpha_1)U_{\rm stagg}$, the definition of the reflection map
	$\vartheta$ and (\ref{Ualpha1tildeHK1}), we obtain the desired expression,
	\begin{equation}
		\label{hatHK10}
		\begin{split}
			\hat{H}_{{\rm K,u},1}^0:=(\tilde{U}_1)^\dagger\tilde{H}_{{\rm K,u},1}^0\tilde{U}_1&=
			\frac{i\kappa}{2}\sum_{x\in\Lambda\; :\; x^{(1)}=0}\sum_{i=1}^4[\xi_{\rm u}^{(i)}(x)\xi_{\rm u}^{(i)}(x+e_1)
			-\eta_{\rm u}^{(i)}(x)\eta_{\rm u}^{(i)}(x+e_1)] \\
			&\quad+\frac{i\kappa}{2}\sum_{x\in\Lambda\; :\: x^{(1)}=L}\sum_{i=1}^4[\xi_{\rm u}^{(i)}(x_L^-)\xi_{\rm u}^{(i)}(x)
			-\eta_{\rm u}^{(i)}(x_L^-)\eta_{\rm u}^{(i)}(x)] \\
			&=\frac{i\kappa}{2}\sum_{x\in\Lambda\; :\; x^{(1)}=0}\sum_{i=1}^4[\xi_{\rm u}^{(i)}(x)\vartheta(\xi_{\rm u}^{(i)}(x))
			+\eta_{\rm u}^{(i)}(x)\vartheta(\eta_{\rm u}^{(i)}(x))] \\
			&\quad+\frac{i\kappa}{2}\sum_{x\in\Lambda\; :\: x^{(1)}=-L+1}\sum_{i=1}^4[\xi_{\rm u}^{(i)}(x)\vartheta(\xi_{\rm u}^{(i)}(x))
			+\eta_{\rm u}^{(i)}(x)\vartheta(\eta_{\rm u}^{(i)}(x))].
		\end{split}
	\end{equation}
	This is nothing but the desired form for the reflection positivity \cite{JP,GK}.

	As to $\tilde{H}_{{\rm K,u},1}^\pm$ of (\ref{tildeHK1+}) and (\ref{tildeHK1-}), one notices that
	both of the two Hamiltonians do not change under the $U(\alpha_1)$ transformation of (\ref{Ualpha1}).
	Further, since the matrix $\alpha_1$ is symmetric, i.e., its transpose equals itself, one has
	\begin{align*}
		& (U_{\rm odd,u})^\dagger[\Psi_{\rm u}^\dagger(x)\alpha_1\Psi_{\rm u}(x+e_1)
		-\Psi_{\rm u}^\dagger(x+e_1)\alpha_1\Psi_{\rm u}(x)]U_{\rm odd,u}\\
		&=\Psi_{\rm u}^\dagger(x)\alpha_1\T\Psi_{\rm u}^\dagger(x+e_1)-\T\Psi_{\rm u}(x+e_1)\alpha_1\Psi_{\rm u}(x)
	\end{align*}
	for any $x\in\Lambda$, where the superscript `$\mathrm{t}$' denotes the transpose, namely
	\[
	\T\Psi^\dagger_{\rm f}(x)=
	\begin{pmatrix}
		\psi_{\rm f}^{(1)}(x)^\dagger \\ \psi_{\rm f}^{(2)}(x)^\dagger \\ \psi_{\rm f}^{(3)}(x)^\dagger \\ \psi_{\rm f}^{(4)}(x)^\dagger
	\end{pmatrix},
	\quad
	\T\Psi_{\rm f}(x)=\left(\psi_{\rm f}^{(1)}(x), \psi_{\rm f}^{(2)}(x), \psi_{\rm f}^{(3)}(x), \psi_{\rm f}^{(4)}(x) \right).
	\]
	Therefore, we have
	\begin{equation*}
		\hat{H}_{{\rm K,u},1}^+:=(\tilde{U}_1)^\dagger\tilde{H}_{{\rm K, u},1}^+\tilde{U}_1
		=-i\kappa\sum_{x\in\Lambda_+\; : \; x^{(1)}\ne L}[\Psi_{\rm u}^\dagger(x)\alpha_1\T\Psi_{\rm u}^\dagger(x+e_1)-\T\Psi_{\rm u}(x+e_1)\alpha_1\Psi_{\rm u}(x)]
	\end{equation*}
	and
	\begin{equation*}
		\hat{H}_{{\rm K,u},1}^-:=(\tilde{U}_1)^\dagger\tilde{H}_{{\rm K,u},1}^-\tilde{U}_1
		=-i\kappa\sum_{x\in\Lambda_-\; : \; x^{(1)}\ne 0}[\Psi_{\rm u}^\dagger(x)\alpha_1\T\Psi_{\rm u}^\dagger(x+e_1)-\T\Psi_{\rm u}(x+e_1)\alpha_1\Psi_{\rm u}(x)].
	\end{equation*}
	In addition, since the reflection map $\vartheta$ changes the hopping direction, we obtain
	\begin{equation*}
		\hat{H}_{{\rm K,u},1}^+=\vartheta(\hat{H}_{{\rm K,u},1}^-),
	\end{equation*}
	where we have used that the matrix $\alpha_1$ is real hermitian. In passing, since this argument is independent of the sign of
	the hopping amplitude, the relation holds in the case of the flavor d.

	Next, consider the kinetic Hamiltonian $\tilde{H}_{{\rm K,u},3}^\pm$ of (\ref{tildeHK3pm}) in the third direction.
	From the properties of the matrices $\alpha_i$, one notices that
	\begin{equation*}
		(U(\alpha_1))^\dagger \tilde{H}_{{\rm K,u},3}^+U(\alpha_1)=-\tilde{H}_{{\rm K,u},3}^+
		\quad \mbox{and} \quad
		(U(\alpha_1))^\dagger \tilde{H}_{{\rm K,u},3}^-U(\alpha_1)=\tilde{H}_{{\rm K,u},3}^-.
	\end{equation*}
	Further, one has
	\begin{align*}
		& (U_{\rm odd,u})^\dagger[\Psi_{\rm u}^\dagger(x)\alpha_3\Psi_{\rm u}(x+e_3)-\Psi_{\rm u}^\dagger(x+e_3)\alpha_3\Psi_{\rm u}(x)]U_{\rm odd,u}\\
		&=\Psi_{\rm u}^\dagger(x)\alpha_3\T\Psi_{\rm u}^\dagger(x+e_3)-\T\Psi_{\rm u}(x+e_3)\alpha_3\Psi_{\rm u}(x)
	\end{align*}
	because the matrix $\alpha_3$ is symmetric. From these observations, we have
	\begin{equation}
		\label{hatHK3pm}
		\hat{H}_{{\rm K,u},3}^\pm :=(\tilde{U}_1)^\dagger\tilde{H}_{{\rm K,u},3}^\pm \tilde{U}_1
		=\pm i\kappa\sum_{x\in\Lambda_\pm}[\Psi_{\rm u}^\dagger(x)\alpha_3\T\Psi_{\rm u}^\dagger(x+e_3)
		-\T\Psi_{\rm u}(x+e_3)\alpha_3\Psi_{\rm u}(x)].
	\end{equation}
	This implies
	\begin{equation*}
		\hat{H}_{{\rm K,u},3}^+=\vartheta(\hat{H}_{{\rm K,u},3}^-)
	\end{equation*}
	because the matrix $\alpha_3$ is real hermitian.

	Finally, let us consider $\tilde{H}_{{\rm K,u},2}^\pm$ of (\ref{tildeHK2pm}). In the same way, one has
	\begin{equation*}
		[U(\alpha_1)]^\dagger\tilde{H}_{{\rm K,u},2}^\pm U(\alpha_1)
		=\mp \kappa\sum_{x\in\Lambda_\pm}[\Psi_{\rm u}^\dagger(x)\alpha_2\Psi_{\rm u}(x+e_2)+\Psi_{\rm u}^\dagger(x+e_2)\alpha_2\Psi_{\rm u}(x)].
	\end{equation*}
	Note that
	\begin{align*}
		&(U_{\rm odd,u})^\dagger[\Psi_{\rm u}^\dagger(x)\alpha_2\Psi_{\rm u}(x+e_2)
		+\Psi_{\rm u}^\dagger(x+e_2)\alpha_2\Psi_{\rm u}(x)]U_{\rm odd,u} \\
		&=\Psi_{\rm u}^\dagger(x)\alpha_2\T\Psi_{\rm u}^\dagger(x+e_2)+\T\Psi_{\rm u}(x+e_2)\alpha_2\Psi_{\rm u}(x)
	\end{align*}
	for any $x\in\Lambda$, where we have used that the matrix $\alpha_2$ is anti-symmetric.
	By combining these two, we obtain
	\begin{equation*}
		\hat{H}_{{\rm K,u},2}^\pm:=(\tilde{U}_1)^\dagger\tilde{H}_{{\rm K,u},2}^\pm \tilde{U}_1
		=\mp \kappa\sum_{x\in\Lambda_\pm}[\Psi_{\rm u}^\dagger(x)\alpha_2\T\Psi_{\rm u}^\dagger(x+e_2)+\T\Psi_{\rm u}(x+e_2)\alpha_2\Psi_{\rm u}(x)].
	\end{equation*}
	Since the matrix $\alpha_2$ is pure imaginary hermitian, we have
	\begin{equation*}
		\hat{H}_{{\rm K,u},2}^+=\vartheta(\hat{H}_{{\rm K,u},2}^-).
	\end{equation*}

	As mentioned in some comments above, these arguments hold for the case of the flavor d.

	\subsection{Interaction Hamiltonian $\tilde{H}_{\rm int,R}^{(\Lambda)}(h)$}

	Next, consider the interaction Hamiltonian $\tilde{H}_{\rm int,R}^{(\Lambda)}(h)$ of (\ref{tildeHintRhf}).
	For the part about the operator $\tilde{\Gamma}^{(1)}(x)$, we write
	\begin{equation}
		\label{tildeHtildeGamma1}
		\tilde{H}_{\tilde{\Gamma}^{(1)}}^{(\Lambda)}(h)=\sum_{\mu=1}^3\tilde{H}_{\tilde{\Gamma}^{(1)},\mu}^{(\Lambda)}(h)
		-3g\sum_{x\in\Lambda}[\tilde{\Gamma}^{(1)}(x)]^2,
	\end{equation}
	where
	\begin{equation*}
		\tilde{H}_{\tilde{\Gamma}^{(1)},\mu}^{(\Lambda)}(h):=\frac{g}{2}\sum_{x\in\Lambda}
		[\tilde{\Gamma}^{(1)}(x)-\tilde{\Gamma}^{(1)}(x+e_\mu)+h_\mu(x)]^2.
	\end{equation*}
	The Hamiltonian $\tilde{H}_{\tilde{\Gamma}^{(1)},1}^{(\Lambda)}(h)$ can be decomposed into three parts as follows:
	\begin{equation*}
		\tilde{H}_{\tilde{\Gamma}^{(1)},1}^{(\Lambda)}(h)
		=\tilde{H}_{\tilde{\Gamma}^{(1)},1}^+(h)+\tilde{H}_{\tilde{\Gamma}^{(1)},1}^-(h)
		+\tilde{H}_{\tilde{\Gamma}^{(1)},1}^0(h),
	\end{equation*}
	where
	\begin{equation*}
		\tilde{H}_{\tilde{\Gamma}^{(1)},1}^+(h):=\frac{g}{2}\sum_{\substack{x\in\Lambda_+:\\ x^{(1)}\ne L}}
		[\tilde{\Gamma}^{(1)}(x)-\tilde{\Gamma}^{(1)}(x+e_1)+h_1(x)]^2,
	\end{equation*}
	\begin{equation*}
		\tilde{H}_{\tilde{\Gamma}^{(1)},1}^-(h):=\frac{g}{2}\sum_{\substack{x\in\Lambda_-:\\ x^{(1)}\ne 0}}
		[\tilde{\Gamma}^{(1)}(x)-\tilde{\Gamma}^{(1)}(x+e_1)+h_1(x)]^2,
	\end{equation*}
	and
	\begin{equation}
		\label{Hint10h}
		\tilde{H}_{\tilde{\Gamma}^{(1)},1}^0(h):=\frac{g}{2}\sum_{\substack{x\in\Lambda:\\ x^{(1)}=0,L}}
		[\tilde{\Gamma}^{(1)}(x)-\tilde{\Gamma}^{(1)}(x+e_1)+h_1(x)]^2.
	\end{equation}
	Similarly, we have
	\begin{equation*}
		\tilde{H}_{\tilde{\Gamma}^{(1)},\mu}^{(\Lambda)}(h)
		=\tilde{H}_{\tilde{\Gamma}^{(1)},\mu}^+(h)+\tilde{H}_{\tilde{\Gamma}^{(1)},\mu}^-(h) \ \ \mbox{for \ } \mu=2,3,
	\end{equation*}
	where
	\begin{equation*}
		\tilde{H}_{\tilde{\Gamma}^{(1)},\mu}^\pm(h)
		:=\frac{g}{2}\sum_{x\in\Lambda_\pm}[\tilde{\Gamma}^{(1)}(x)-\tilde{\Gamma}^{(1)}(x+e_\mu)+h_\mu(x)]^2\ \mbox{for \ }\mu=2,3.
	\end{equation*}
	From the definitions of $\tilde{\Gamma}^{(1)}(x)$ and $U(\alpha_1)$, one has
	\begin{equation*}
		[U(\alpha_1)]^\dagger \tilde{\Gamma}^{(1)}(x)U(\alpha_1)=\begin{cases}
			-\Psi_{\rm u}^\dagger(x)\gamma_0\Psi_{\rm u}(x)+\Psi_{\rm d}^\dagger(x)\gamma_0\Psi_{\rm d}(x) & \mbox{if \ } x\in\Lambda_+;\\
			\Psi_{\rm u}^\dagger(x)\gamma_0\Psi_{\rm u}(x)-\Psi_{\rm d}^\dagger(x)\gamma_0\Psi_{\rm d}(x) & \mbox{if \ } x\in\Lambda_-.
		\end{cases}
	\end{equation*}
	Further, by using $\tilde{U}_1:=U(\alpha_1)U_{\rm stagg}$, we have
	\begin{equation}
		\label{tildeU1tildeGamma1}
		(\tilde{U}_1)^\dagger \tilde{\Gamma}^{(1)}(x)\tilde{U}_1=\mp (-1)^{x^{(1)}+x^{(2)}+x^{(3)}}\tilde{\Gamma}^{(1)}(x)\quad \mbox{for \ }
		x\in\Lambda_\pm,
	\end{equation}
	where we have used the expression of the matrix $\gamma_0$ in (\ref{gamma05}).
	For the Hamiltonian $\tilde{H}_{\tilde{\Gamma}^{(1)},1}^0(h)$ of (\ref{Hint10h}), we have
	\begin{align*}
		\tilde{U}_1^\dagger \tilde{H}_{\tilde{\Gamma}^{(1)},1}^0(h)\tilde{U}_1
		=&\frac{g}{2}\sum_{\substack{x\in\Lambda:\\ x^{(1)}=0}}
		[\tilde{\Gamma}^{(1)}(x)-\tilde{\Gamma}^{(1)}(x+e_1)+(-1)^{x^{(1)}+x^{(2)}+x^{(3)}}h_1(x)]^2 \\
		&+\frac{g}{2}\sum_{\substack{x\in\Lambda:\\ x^{(1)}=L}}
		[\tilde{\Gamma}^{(1)}(x)-\tilde{\Gamma}^{(1)}(x+e_1)-(-1)^{x^{(1)}+x^{(2)}+x^{(3)}}h_1(x)]^2. \\
	\end{align*}
	This is the desired form \cite{DLS,GK} for getting a Gaussian domination.
	Similarly, one has
	\begin{equation*}
		\tilde{U}_1^\dagger \tilde{H}_{\tilde{\Gamma}^{(1)},1}^+(h)\tilde{U}_1
		=\frac{g}{2}\sum_{\substack{x\in\Lambda_+:\\ x^{(1)}\ne L}} [\tilde{\Gamma}^{(1)}(x)+\tilde{\Gamma}^{(1)}(x+e_1)-\tilde{h}_1(x)]^2
	\end{equation*}
	with $\tilde{h}_1(x):=(-1)^{x^{(1)}+x^{(2)}+x^{(3)}}h_1(x)$, and
	\begin{equation*}
		\tilde{U}_1^\dagger \tilde{H}_{\tilde{\Gamma}^{(1)},1}^-(h)\tilde{U}_1
		=\frac{g}{2}\sum_{\substack{x\in\Lambda_-:\\ x^{(1)}\ne 0}} [\tilde{\Gamma}^{(1)}(x)+\tilde{\Gamma}^{(1)}(x+e_1)+\tilde{h}_1(x)]^2.
	\end{equation*}
	Moreover for $\mu=2,3$, we have
	\begin{equation*}
		\tilde{U}_1^\dagger \tilde{H}_{\tilde{\Gamma}^{(1)},\mu}^+(h)\tilde{U}_1
		=\frac{g}{2}\sum_{x\in\Lambda_+}[\tilde{\Gamma}^{(1)}(x)+\tilde{\Gamma}^{(1)}(x+e_\mu)-\tilde{h}_\mu(x)]^2
	\end{equation*}
	and
	\begin{equation*}
		\tilde{U}_1^\dagger \tilde{H}_{\tilde{\Gamma}^{(1)},\mu}^-(h)\tilde{U}_1
		=\frac{g}{2}\sum_{x\in\Lambda_-}[\tilde{\Gamma}^{(1)}(x)+\tilde{\Gamma}^{(1)}(x+e_\mu)+\tilde{h}_\mu(x)]^2
	\end{equation*}
	with $\tilde{h}_\mu(x):=(-1)^{x^{(1)}+x^{(2)}+x^{(3)}}h_\mu(x)$. In particular, when $h=0$, these imply
	\begin{equation*}
		\vartheta(\hat{H}_{\tilde{\Gamma}^{(1)},\mu}^-(0))=
		\hat{H}_{\tilde{\Gamma}^{(1)},\mu}^+(0)\quad \mbox{for \ } \mu=1,2,3,
	\end{equation*}
	where we have written
	\begin{equation*}
		\hat{H}_{\tilde{\Gamma}^{(1)},\mu}^\pm(h):=\tilde{U}_1^\dagger \tilde{H}_{\tilde{\Gamma}^{(1)},\mu}^\pm(h)\tilde{U}_1
	\end{equation*}
	for $\mu=1,2,3$.

	Clearly, the second sum in the right hand side of (\ref{tildeHtildeGamma1}) can be treated in the same way.

	In the above argument, the relation (\ref{tildeU1tildeGamma1}) is crucial for the reflection positivity.
	Therefore, as to the three operators, $\tilde{S}^{(1)}(x)$, $\tilde{S}^{(3)}(x)$ and $\tilde{S}_5^{(2)}(x)$, in
	the Hamiltonian $\tilde{H}_{\rm int,R}^{(\Lambda)}(h)$,
	it is enough to check the corresponding relations. Actually, in the same way, we have
	\begin{equation*}
		\tilde{U}_1^\dagger \tilde{S}^{(1)}(x)\tilde{U}_1=\pm (-1)^{x^{(1)}+x^{(2)}+x^{(3)}}[\T\Psi_{\rm u}(x)\alpha_1\Psi_{\rm d}(x)
		+\Psi_{\rm d}^\dagger(x)\alpha_1\T\Psi_{\rm u}^\dagger(x)]\ \ \mbox{for \ } x\in\Lambda_\pm,
	\end{equation*}
	\begin{equation*}
		\tilde{U}_1^\dagger \tilde{S}^{(3)}(x)\tilde{U}_1=\pm (-1)^{x^{(1)}+x^{(2)}+x^{(3)}}[-\Psi_{\rm u}^\dagger(x)\gamma_0\Psi_{\rm u}(x)
		+\Psi_{\rm d}^\dagger(x)\gamma_0\Psi_{\rm d}(x)]\ \ \mbox{for \ } x\in\Lambda_\pm
	\end{equation*}
	and
	\begin{equation*}
		\tilde{U}_1^\dagger \tilde{S}_5^{(2)}(x)\tilde{U}_1=
		\mp (-1)^{x^{(1)}+x^{(2)}+x^{(3)}}
		[\Psi_{\rm u}^\dagger(x)\gamma_5\alpha_1\T\Psi_{\rm d}^\dagger(x)+\T\Psi_{\rm d}(x)\gamma_5\alpha_1\Psi_{\rm u}(x)]
		\ \ \mbox{for \ } x\in\Lambda_\pm.
	\end{equation*}

	\subsection{Interaction Hamiltonian $\tilde{H}_{\rm int,I}^{(\Lambda)}$}

	Let us consider the interaction Hamiltonian $\tilde{H}_{{\rm int},{\rm I}}^{(\Lambda)}$ of (\ref{tildeHintI}).
	The part about the operator $\tilde{\Gamma}^{(2)}(x)$ of (\ref{tildeGamma2}) is given by
	\begin{equation}
		\label{tildeHtildeGamma2}
		\tilde{H}_{\tilde{\Gamma}^{(2)}}^{(\Lambda)}
		:=\sum_{\mu=1}^3\tilde{H}_{\tilde{\Gamma}^{(2)},\mu}^{(\Lambda)}+3g\sum_{x\in\Lambda}[\tilde{\Gamma}^{(2)}(x)]^2,
	\end{equation}
	where
	\begin{equation*}
		\tilde{H}_{\tilde{\Gamma}^{(2)},\mu}^{(\Lambda)}
		:=-\frac{g}{2}\sum_{x\in\Lambda}[\tilde{\Gamma}^{(2)}(x)+\tilde{\Gamma}^{(2)}(x+e_\mu)]^2.
	\end{equation*}
	The Hamiltonian $\tilde{H}_{\tilde{\Gamma}^{(2)},1}^{(\Lambda)}$ can be decomposed into three parts as follows:
	\begin{equation*}
		\tilde{H}_{\tilde{\Gamma}^{(2)},1}^{(\Lambda)}
		=\tilde{H}_{\tilde{\Gamma}^{(2)},1}^++\tilde{H}_{\tilde{\Gamma}^{(2)},1}^-
		+\tilde{H}_{\tilde{\Gamma}^{(2)},1}^0,
	\end{equation*}
	where
	\begin{equation*}
		\tilde{H}_{\tilde{\Gamma}^{(2)},1}^+:=-\frac{g}{2}\sum_{\substack{x\in\Lambda_+:\\ x^{(1)}\ne L}}
		[\tilde{\Gamma}^{(2)}(x)+\tilde{\Gamma}^{(2)}(x+e_1)]^2,
	\end{equation*}
	\begin{equation*}
		\tilde{H}_{\tilde{\Gamma}^{(2)},1}^-:=-\frac{g}{2}\sum_{\substack{x\in\Lambda_-:\\ x^{(1)}\ne 0}}
		[\tilde{\Gamma}^{(2)}(x)+\tilde{\Gamma}^{(2)}(x+e_1)]^2,
	\end{equation*}
	and
	\begin{equation}
		\label{tildeHGamma210}
		\tilde{H}_{\tilde{\Gamma}^{(2)},1}^0:=-\frac{g}{2}\sum_{\substack{x\in\Lambda:\\ x^{(1)}=0,L}}
		[\tilde{\Gamma}^{(2)}(x)+\tilde{\Gamma}^{(2)}(x+e_1)]^2.
	\end{equation}
	Similarly, we have
	\begin{equation*}
		\tilde{H}_{\tilde{\Gamma}^{(2)},\mu}^{(\Lambda)}
		=\tilde{H}_{\tilde{\Gamma}^{(2)},\mu}^++\tilde{H}_{\tilde{\Gamma}^{(2)},\mu}^- \ \ \mbox{for \ } \mu=2,3,
	\end{equation*}
	where
	\begin{equation*}
		\tilde{H}_{\tilde{\Gamma}^{(2)},\mu}^\pm
		:=-\frac{g}{2}\sum_{x\in\Lambda_\pm}[\tilde{\Gamma}^{(2)}(x)+\tilde{\Gamma}^{(2)}(x+e_\mu)]^2\ \ \mbox{for \ }\mu=2,3.
	\end{equation*}

	One has
	\begin{equation*}
		[U(\alpha_1)]^\dagger \tilde{\Gamma}^{(2)}(x)U(\alpha_1)=\begin{cases}
			\Psi_{\rm u}^\dagger(x)i\gamma_0\gamma_5\Psi_{\rm u}(x)+\Psi_{\rm d}^\dagger(x)i\gamma_0\gamma_5\Psi_{\rm d}(x) & \mbox{if \ } x\in\Lambda_+;\\
			-\Psi_{\rm u}^\dagger(x)i\gamma_0\gamma_5\Psi_{\rm u}(x)-\Psi_{\rm d}^\dagger(x)i\gamma_0\gamma_5\Psi_{\rm d}(x) & \mbox{if \ } x\in\Lambda_-.
		\end{cases}
	\end{equation*}
	Therefore, we obtain
	\begin{equation}
		\label{tildeU1Gamma2}
		(\tilde{U}_1)^\dagger \tilde{\Gamma}^{(2)}(x)\tilde{U}_1=\pm \Gamma^{(2)}(x)\quad \mbox{for \ } x\in\Lambda_\pm,
	\end{equation}
	where we have used that the matrix $\gamma_0\gamma_5$ is anti-symmetric.
	For the Hamiltonian $\tilde{H}_{\tilde{\Gamma}^{(2)},1}^0$ of (\ref{tildeHGamma210}), we have
	\begin{equation*}
		\tilde{U}_1^\dagger \tilde{H}_{\tilde{\Gamma}^{(2)},1}^0\tilde{U}_1
		=-\frac{g}{2}\sum_{\substack{x\in\Lambda: \\ x^{(1)}=0,L}}[\Gamma^{(2)}(x)-\Gamma^{(2)}(x+e_1)]^2.
	\end{equation*}
	This is the desired form \cite{DLS,GK} for getting a Gaussian domination. Similarly, one has
	\begin{equation*}
		\tilde{U}_1^\dagger \tilde{H}_{\tilde{\Gamma}^{(2)},1}^+\tilde{U}_1
		=-\frac{g}{2}\sum_{\substack{x\in\Lambda_+: \\ x^{(1)}\ne L}}[\Gamma^{(2)}(x)+\Gamma^{(2)}(x+e_1)]^2
	\end{equation*}
	and
	\begin{equation*}
		\tilde{U}_1^\dagger \tilde{H}_{\tilde{\Gamma}^{(2)},1}^-\tilde{U}_1
		=-\frac{g}{2}\sum_{\substack{x\in\Lambda_-: \\ x^{(1)}\ne 0}}[\Gamma^{(2)}(x)+\Gamma^{(2)}(x+e_1)]^2.
	\end{equation*}
	Further, we have
	\begin{equation*}
		\tilde{U}_1^\dagger \tilde{H}_{\tilde{\Gamma}^{(2)},\mu}^+\tilde{U}_1
		=-\frac{g}{2}\sum_{x\in\Lambda_+}[\Gamma^{(2)}(x)+\Gamma^{(2)}(x+e_\mu)]^2,
	\end{equation*}
	and
	\begin{equation*}
		\tilde{U}_1^\dagger \tilde{H}_{\tilde{\Gamma}^{(2)},\mu}^-\tilde{U}_1
		=-\frac{g}{2}\sum_{x\in\Lambda_-}[\Gamma^{(2)}(x)+\Gamma^{(2)}(x+e_\mu)]^2
	\end{equation*}
	for $\mu=2,3$. From these observations, we have
	\begin{equation*}
		\vartheta\left(\hat{H}_{\tilde{\Gamma}^{(2)},\mu}^-\right)=\hat{H}_{\tilde{\Gamma}^{(2)},\mu}^+\quad \mbox{for \ } \mu=1,2,3,
	\end{equation*}
	where we have written
	\begin{equation*}
		\hat{H}_{\tilde{\Gamma}^{(2)},\mu}^\pm:=\tilde{U}_1^\dagger \tilde{H}_{\tilde{\Gamma}^{(2)},\mu}^\pm\tilde{U}_1
		\quad \mbox{for \ } \mu=1,2,3,
	\end{equation*}
	and we have used that the matrix $i\gamma_0\gamma_5$ in $\Gamma^{(2)}(x)$ is pure imaginary hermitian.

	Clearly, the second sum in the right-hand side of (\ref{tildeHtildeGamma2}) can be treated in the same way.

	In the above argument, the relation (\ref{tildeU1Gamma2}) is crucial again.
	Therefore, as to the rest of the operators in the interaction Hamiltonian $\tilde{H}_{{\rm int},{\rm I}}^{(\Lambda)}$
	of (\ref{tildeHintI}), it is enough to check the corresponding relations.
	One can check that
	\begin{equation*}
		\tilde{U}_1^\dagger \tilde{S}^{(2)}(x)\tilde{U}_1=\pm i[\Psi_{\rm u}^\dagger(x)\alpha_1\T\Psi_{\rm d}^\dagger(x)
		-\T\Psi_{\rm d}(x)\alpha_1\Psi_{\rm u}(x)]\ \ \mbox{for \ } x\in\Lambda_\pm,
	\end{equation*}
	\begin{equation*}
		\tilde{U}_1^\dagger \tilde{S}_5^{(1)}(x)\tilde{U}_1=\mp i[\Psi_{\rm u}^\dagger(x)\alpha_1\gamma_5\T\Psi_{\rm d}^\dagger(x)
		-\T\Psi_{\rm d}(x)\alpha_1\gamma_5\Psi_{\rm u}(x)]\ \ \mbox{for \ } x\in\Lambda_\pm
	\end{equation*}
	and
	\begin{equation*}
		\tilde{U}_1^\dagger \tilde{S}_5^{(3)}(x)\tilde{U}_1=\pm S_5^{(3)}(x) \ \ \mbox{for \ } x\in\Lambda_\pm.
	\end{equation*}

	\section{Reflection with respect to the  plane $x^{(2)}=1/2$}
	\label{sec:x2plane}

	As to the reflection with respect to the {$x^{(3)}=1/2$ }plane, the argument is the same as in the above case of the $x^{(1)}=1/2$ plane.
	Therefore, it is enough to deal with the case of the reflection with respect to the $x^{(2)}=1/2$ plane.

	We write
	\begin{equation*}
		u_3(\theta):=\exp\left[i\frac{\theta}{2}\sigma_3\right]
	\end{equation*}
	for the spin rotation by the angle $\theta\in[0,2\pi)$ about the third axis, and
	\begin{equation*}
		\mathcal{U}_3(\theta):=\begin{pmatrix}
			0 & u_3(\theta) \\
			u_3(\theta) & 0
		\end{pmatrix}
	\end{equation*}
	for the four-component Dirac spinor. We also write $U_3(\theta)$ for the corresponding unitary operator on the fermion
	Fock space, i.e.,
	\begin{equation*}
		[U_3(\theta)]^\dagger\Psi(x)U_3(\theta)=\mathcal{U}_3(\theta)\Psi(x)\quad \mbox{for \ } x\in\Lambda.
	\end{equation*}
	Note that
	\begin{equation*}
		[\mathcal{U}_3(\theta)]^\dagger\alpha_i\mathcal{U}_3(\theta)
		=\begin{pmatrix}
			0 & [u_3(\theta)]^\dagger \sigma_i u_3(\theta) \\
			[u_3(\theta)]^\dagger \sigma_i u_3(\theta) & 0
		\end{pmatrix}
		\quad \mbox{for \ } i=1,2,3,
	\end{equation*}
	\begin{equation*}
		[u_3(\theta)]^\dagger\sigma_1u_3(\theta)=\sigma_1\cos \theta +\sigma_2\sin\theta ,
	\end{equation*}
	\begin{equation*}
		[u_3(\theta)]^\dagger\sigma_2u_3(\theta)=\sigma_2\cos \theta -\sigma_1\sin\theta ,
	\end{equation*}
	and
	\begin{equation*}
		[u_3(\theta)]^\dagger\sigma_3u_3(\theta)=\sigma_3.
	\end{equation*}
	Therefore, one has
	\begin{equation}
		\label{rotalpha1}
		[\mathcal{U}_3(-\pi/2)]^\dagger\alpha_1\mathcal{U}_3(-\pi/2)=-\alpha_2,
	\end{equation}
	\begin{equation}
		\label{rotalpha2}
		[\mathcal{U}_3(-\pi/2)]^\dagger\alpha_2\mathcal{U}_3(-\pi/2)=\alpha_1,
	\end{equation}
	and
	\begin{equation}
		\label{rotalpha3}
		[\mathcal{U}_3(-\pi/2)]^\dagger\alpha_3\mathcal{U}_3(-\pi/2)=\alpha_3.
	\end{equation}
	Moreover,
	\begin{equation}
		\label{rotgamma0}
		[\mathcal{U}_3(-\pi/2)]^\dagger\gamma_0\mathcal{U}_3(-\pi/2)=-\gamma_0
	\end{equation}
	and
	\begin{equation}
		\label{rotgamma5}
		[\mathcal{U}_3(-\pi/2)]^\dagger\gamma_5\mathcal{U}_3(-\pi/2)=\gamma_5.
	\end{equation}
	Clearly, these change only the matrices, $\alpha_1$ and $\alpha_2$, of
	the hopping terms in the $x^{(1)}$ and $x^{(2)}$ directions in the kinetic Hamiltonian.
	In addition, although the coefficient $\kappa$ in front of the matrix $\alpha_2$ changes its sign,
	it does not affect the above argument about the reflection positivity in the case of the $x^{(1)}=1/2$ plane.

	\section{Gaussian domination and infrared bound}
	\label{GdomiInfrab}

	By reflection positivity, we obtain the Gaussian domination bound
	\begin{equation*}
		{\rm Tr}\exp[-\beta H^{(\Lambda)}(h)]\le {\rm Tr}\exp[-\beta H^{(\Lambda)}(0)],
	\end{equation*}
	for the Hamiltonian $H^{(\Lambda)}(h)$ of (\ref{Hmhf}).
	The proof proceeds in a similar manner to that in~\cite{GK, GK2}, and is omitted here to avoid repetition.
	In order to expand this with respect to the function $h$, we note that
	\begin{align*}
		&\frac{g}{2}\sum_{x\in\Lambda}\sum_{\mu=1}^3[\Gamma^{(1)}(x)-\Gamma^{(1)}(x+e_\mu)+h_\mu(x)]^2\\
		&=\frac{g}{2}\sum_{x\in\Lambda}\sum_{\mu=1}^3[\Gamma^{(1)}(x)-\Gamma^{(1)}(x+e_\mu)]^2\\
		&\quad+g\sum_{x\in\Lambda}\sum_{\mu=1}^3[\Gamma^{(1)}(x)-\Gamma^{(1)}(x+e_\mu)]h_\mu(x)
		+\frac{g}{2}\sum_{x\in\Lambda}\sum_{\mu=1}^3[h_\mu(x)]^2.
	\end{align*}
	The second sum in the right-hand side is written
	\begin{equation*}
		g\sum_{x\in\Lambda}\sum_{\mu=1}^3[\Gamma^{(1)}(x)-\Gamma^{(1)}(x+e_\mu)]h_\mu(x)
		=g\sum_{x\in\Lambda}\sum_{\mu=1}^3\Gamma^{(1)}(x)\partial_\mu h_\mu(x),
	\end{equation*}
	where we have written
	\begin{equation*}
		\partial_\mu h_\mu(x):=h_\mu(x)-h_\mu(x-e_\mu).
	\end{equation*}
	Therefore, the second order of the expansion yields
	\begin{equation*}
		(\beta g)^2\Bigl(\sum_{x\in\Lambda}\sum_{\mu=1}^3\Gamma^{(1)}(x)\partial_\mu h_\mu(x),
		\sum_{y\in\Lambda}\sum_{\mu'=1}\Gamma^{(1)}(y)\partial_{\mu'}h_{\mu'}(y)\Bigr)_{\beta}^{(\Lambda)}
		-\frac{\beta g}{2}\sum_{x\in\Lambda}\sum_{\mu=1}^3[h_\mu(x)]^2\le 0,
	\end{equation*}
	where $(\cdots,\cdots)_{\beta}^{(\Lambda)}$ denotes the Duhamel two-point function \cite{DLS}.
	Since this inequality can be extended to \cite{DLS} complex-valued functions $h_\mu(x)$, we have
	\begin{equation}
		\label{Duhambound}
		\Bigl(\sum_{x\in\Lambda}\sum_{\mu=1}^3\Gamma^{(1)}(x)\overline{\partial_\mu h_\mu(x)},
		\sum_{y\in\Lambda}\sum_{\mu'=1}^3\Gamma^{(1)}(y)\partial_{\mu'} h_{\mu'}(y)\Bigr)_{\beta}^{(\Lambda)}
		\le \frac{1}{2\beta g}\sum_{x\in\Lambda}\sum_{\mu=1}^3|h_\mu(x)|^2,
	\end{equation}
	where $\overline{\cdots}$ denotes its complex conjugate. We choose
	\begin{equation*}
		h_\mu(x)=\frac{1}{\sqrt{|\Lambda|}}\bigl[e^{ip(x+e_\mu)}-e^{ipx}\bigr]
	\end{equation*}
	with the wavenumber $p=(p^{(1)},p^{(2)},p^{(3)})$, where we have written $px=p^{(1)}x^{(1)}+p^{(2)}x^{(2)}+p^{(3)}x^{(3)}$.
	Then, one has
	\begin{equation*}
		\partial_\mu h_\mu(x)=-\frac{2}{\sqrt{|\Lambda|}}e^{ipx}(1-\cos p^{(\mu)})
	\end{equation*}
	and
	\begin{equation*}
		\sum_{x\in\Lambda} \sum_{\mu=1}^3 |h_\mu (x)|^2=2E_p,
	\end{equation*}
	where we have written
	\begin{equation*}
		E_p:=\sum_{\mu =1}^3(1-\cos p^{(\mu)}).
	\end{equation*}
	{From} these, we have
	\begin{equation*}
		\sum_{x\in\Lambda}\sum_{\mu=1}^3 \Gamma^{(1)}(x)\partial_\mu h_\mu(x)=-2\hat{\Gamma}_p^{(1)}E_p,
	\end{equation*}
	where we have written
	\begin{equation}
		\label{hatGammap1}
		\hat{\Gamma}_p^{(1)}:=\frac{1}{\sqrt{|\Lambda|}}\sum_{x\in\Lambda}\Gamma^{(1)}(x)e^{ipx}.
	\end{equation}
	By substituting these into the above bound (\ref{Duhambound}), we obtain
	\begin{equation}
		\label{bpbound}
		\mathfrak{b}_p:=(\hat{\Gamma}_{-p}^{(1)},\hat{\Gamma}_p^{(1)})_{\beta}^{(\Lambda)}\le \frac{1}{4\beta gE_p}.
	\end{equation}

	\section{Long-range order}
	\label{sec:LRO}

	Now, let us prove the existence of the long-range order for $\Gamma^{(1)}(x)$. We write
	\begin{equation*}
		\mathfrak{g}_p:=\frac{1}{2}\bigl[\langle \hat{\Gamma}_p^{(1)}\hat{\Gamma}_{-p}^{(1)}\rangle_{\beta}^{(\Lambda)}
		+\langle \hat{\Gamma}_{-p}^{(1)}\hat{\Gamma}_{p}^{(1)}\rangle_{\beta}^{(\Lambda)}\bigr]
	\end{equation*}
	and
	\begin{equation}
		\label{cp}
		\mathfrak{c}_p:=\bigl\langle\bigl[\hat{\Gamma}_{-p}^{(1)},[H^{(\Lambda)}(0),\hat{\Gamma}_p^{(1)}]\bigr]
		\bigr\rangle_{\beta}^{(\Lambda)}
	\end{equation}
	for their thermal expectation values $\langle\cdots\rangle_\beta^{(\Lambda)}$. We use the inequality, \cite{DLS}
	\begin{equation*}
		\mathfrak{g}_p\le \frac{1}{2}\biggl[\mathfrak{b}_p+\sqrt{\mathfrak{b}_p^2+\beta\mathfrak{b}_p\mathfrak{c}_p}\biggr]
		\le \mathfrak{b}_p+\frac{1}{2}\sqrt{\beta\mathfrak{b}_p\mathfrak{c}_p}.
	\end{equation*}
	By using the bound (\ref{bpbound}) for $\mathfrak{b}_p$, we have
	\begin{equation*}
		\frac{1}{2}\bigl[\langle \hat{\Gamma}_p^{(1)}\hat{\Gamma}_{-p}^{(1)}\rangle_{\beta}^{(\Lambda)}
		+\langle \hat{\Gamma}_{-p}^{(1)}\hat{\Gamma}_{p}^{(1)}\rangle_{\beta}^{(\Lambda)}\bigr]
		\le \frac{1}{4\beta gE_p}+\frac{1}{4}\sqrt{\frac{\mathfrak{c}_p}{gE_p}}.
	\end{equation*}
	Further, following \cite{KLS1,KLS2}, we have
	\begin{equation}
		\label{GammaGammacos}
		\frac{1}{3|\Lambda|}\sum_p\langle \hat{\Gamma}_p^{(1)}\hat{\Gamma}_{-p}^{(1)}\rangle_{\beta}^{(\Lambda)}
		\biggl[\sum_{\mu=1}^3\cos p^{(\mu)}\biggr]
		\le \delta(\beta)+\frac{1}{12|\Lambda|}\sum_{p\ne 0}\sqrt{\frac{\mathfrak{c}_p}{gE_p}}
		\biggl\{\sum_{\mu=1}^3\cos p^{(\mu)}\biggr\}_+
		+   \Sigma_{\rm LRO}^{(\Lambda)},
	\end{equation}
	where $\{a\}_+=\max\{a,0\}$, and we have written
	\begin{equation*}
		\delta(\beta):=\frac{1}{12|\Lambda|}\sum_{p\ne 0}\frac{1}{\beta gE_p}
		\biggl\{\sum_{\mu=1}^3\cos p^{(\mu)}\biggr\}_+
	\end{equation*}
	and
	\begin{equation*}
		\Sigma_{\rm LRO}^{(\Lambda)}:=\frac{1}{|\Lambda|}\langle\hat{\Gamma}_0^{(1)}\hat{\Gamma}_0^{(1)} \rangle_{\beta}^{(\Lambda)}
		=\frac{1}{|\Lambda|^2}\sum_{x,y\in\Lambda}\langle \Gamma^{(1)}(x)\Gamma^{(1)}(y)\rangle_{\beta}^{(\Lambda)}.
	\end{equation*}
	Clearly, the quantity $\delta(\beta)$ becomes small for a large $\beta$, and $\Sigma_{\rm LRO}^{(\Lambda)}$ is nothing but
	the long-range order. Our goal is to show that the long-range order $\Sigma_{\rm LRO}^{(\Lambda)}$ is strictly positive
	in the infinite-volume limit $\Lambda\nearrow \mathbb{Z}^3$.

	For this purpose, let us first estimate the second term in the right-hand side of (\ref{GammaGammacos}).
	By using the Schwarz inequality, one has
	\begin{equation}
		\label{sumcp}
		\frac{1}{|\Lambda|}\sum_{p\ne 0}\frac{1}{\sqrt{E_p}}\biggl\{\sum_{\mu=1}^3\cos p^{(\mu)}\biggr\}_+\sqrt{\mathfrak{c}_p}
		\le \sqrt{3}I_3 \sqrt{\frac{1}{|\Lambda|}\sum_{p}\mathfrak{c}_p},
	\end{equation}
	where we have written
	\begin{equation*}
		I_3:=\sqrt{\frac{1}{3|\Lambda|}\sum_{p\ne 0}\frac{1}{E_p} \biggl(\biggl\{\sum_{\mu=1}^3\cos p^{(\mu)}\biggr\}_+\biggr)^2}.
	\end{equation*}

	Let us estimate the right-hand side of (\ref{sumcp}). {From} the definition (\ref{cp}) of $\mathfrak{c}_p$, one has
	\begin{align*}
		\mathfrak{c}_p&=\bigl\langle\bigl[\hat{\Gamma}_{-p}^{(1)},[H^{(\Lambda)}(0),\hat{\Gamma}_p^{(1)}]\bigr]
		\bigr\rangle_{\beta}^{(\Lambda)} \\
		&=\bigl\langle\bigl[\hat{\Gamma}_{-p}^{(1)},[H_{\rm K}^{(\Lambda)},\hat{\Gamma}_p^{(1)}]\bigr]
		\bigr\rangle_{\beta}^{(\Lambda)}+
		\bigl\langle\bigl[\hat{\Gamma}_{-p}^{(1)},[H_{\rm int}^{(\Lambda)},\hat{\Gamma}_p^{(1)}]\bigr]
		\bigr\rangle_{\beta}^{(\Lambda)}.
	\end{align*}
	For the contribution of the kinetic Hamiltonian $H_{\rm K}^{(\Lambda)}$ in the right-hand side, we have
	\begin{align*}
		\frac{1}{|\Lambda|}\sum_p \bigl\langle\bigl[\hat{\Gamma}_{-p}^{(1)},[H_{\rm K}^{(\Lambda)},\hat{\Gamma}_p^{(1)}]\bigr]
		\bigr\rangle_{\beta}^{(\Lambda)}&=\frac{1}{|\Lambda|^2}\sum_p\sum_{x,y\in\Lambda}
		\bigl\langle\bigl[\Gamma^{(1)}(x),[H_{\rm K}^{(\Lambda)},\Gamma^{(1)}(y)]\bigr] \bigr\rangle_{\beta}^{(\Lambda)}
		\; e^{-ip(x-y)} \\
		&=\frac{1}{|\Lambda|}\sum_{x\in\Lambda}
		\bigl\langle\bigl[\Gamma^{(1)}(x),[H_{\rm K}^{(\Lambda)},\Gamma^{(1)}(x)]\bigr] \bigr\rangle_{\beta}^{(\Lambda)}.
	\end{align*}
	Note that
	\begin{align*}
		& [H_{\rm K}^{(\Lambda)},\Gamma^{(1)}(x)] \\
		&=i\kappa\sum_{\mu=1}^3
		\sum_{y=x,x-e_\mu}\{[\Psi^\dagger(y)\alpha_\mu\Psi(y+e_\mu),\Gamma^{(1)}(x)]-[\Psi^\dagger(y+e_\mu)\alpha_\mu\Psi(y),\Gamma^{(1)}(x)]\}.
	\end{align*}
	Clearly, this right-hand side is of order 1. Therefore, we have
	\begin{equation*}
		\left|\frac{1}{|\Lambda|}\sum_p \bigl\langle\bigl[\hat{\Gamma}_{-p}^{(1)},
		[H_{\rm K}^{(\Lambda)},\hat{\Gamma}_p^{(1)}]\bigr]
		\bigr\rangle_{\beta}^{(\Lambda)}\right|\le \mathcal{C}_0|\kappa|,
	\end{equation*}
	where $\mathcal{C}_0$ is a positive constant.

	The contribution for the interaction Hamiltonian $H_{\rm int}^{(\Lambda)}$ is written
	\begin{align*}
		& \frac{1}{|\Lambda|}\sum_p \bigl\langle\bigl[\hat{\Gamma}_{-p}^{(1)},[H_{\rm int}^{(\Lambda)},\hat{\Gamma}_p^{(1)}]\bigr]
		\bigr\rangle_{\beta}^{(\Lambda)} \\
		&=\frac{1}{|\Lambda|^2}\sum_p\sum_{x,y\in\Lambda}
		\bigl\langle\bigl[\Gamma^{(1)}(x),[H_{\rm int}^{(\Lambda)},\Gamma^{(1)}(y)]\bigr] \bigr\rangle_{\beta}^{(\Lambda)}
		\; e^{-ip(x-y)} \\
		&=\frac{1}{|\Lambda|}\sum_{x\in\Lambda}
		\bigl\langle\bigl[\Gamma^{(1)}(x),[H_{\rm int}^{(\Lambda)},\Gamma^{(1)}(x)]\bigr] \bigr\rangle_{\beta}^{(\Lambda)} \\
		&=-\frac{g}{|\Lambda|}\sum_{x\in\Lambda}\sum_{\mu=1}^3\bigl\{
		\langle\bigl[\Gamma^{(1)}(x),[\Gamma^{(2)}(x),\Gamma^{(1)}(x)]\bigr]\Gamma^{(2)}(x+e_\mu)\rangle_{\beta}^{(\Lambda)} \\
		& \qquad \qquad \qquad
		+\langle\bigl[\Gamma^{(1)}(x),[\Gamma^{(2)}(x),\Gamma^{(1)}(x)]\bigr]\Gamma^{(2)}(x-e_\mu)\rangle_{\beta}^{(\Lambda)}\bigr\} \\
		&\quad-\frac{g}{|\Lambda|}\sum_{x\in\Lambda}\sum_{\mu=1}^3\sum_{\ell=1}^3\bigl\{
		\langle\bigl[\Gamma^{(1)}(x),[S_5^{(\ell)}(x),\Gamma^{(1)}(x)]\bigr]S_5^{(\ell)}(x+e_\mu)\rangle_{\beta}^{(\Lambda)} \\
		& \qquad \qquad \qquad
		+\langle\bigl[\Gamma^{(1)}(x),[S_5^{(\ell)}(x),\Gamma^{(1)}(x)]\bigr]S_5^{(\ell)}(x-e_\mu)\rangle_{\beta}^{(\Lambda)}\bigr\},
	\end{align*}
	where we have used the commutation relations, (\ref{commuGamma1S12}) in Appendix~\ref{ChiralRot} for getting the third equality.
	Further, by using the commutation relations (\ref{commuGamma}), (\ref{commuGammaS5}), (\ref{commuGammaS52}),
	and the chiral and the flavor rotational symmetries of the present Hamiltonian, we obtain
	\begin{align*}
		\frac{1}{|\Lambda|}\sum_p \bigl\langle\bigl[\hat{\Gamma}_{-p}^{(1)},[H_{\rm int}^{(\Lambda)},\hat{\Gamma}_p^{(1)}]\bigr]
		\bigr\rangle_{\beta}^{(\Lambda)}&=\frac{8g}{|\Lambda|}\sum_{x\in\Lambda}\sum_{\mu=1}^3
		\langle \Gamma^{(2)}(x)\Gamma^{(2)}(x+e_\mu)\rangle_{\beta}^{(\Lambda)} \\
		&\quad+\frac{8g}{|\Lambda|}\sum_{x\in\Lambda}\sum_{\mu=1}^3\sum_{\ell=1}^3
		\langle S_5^{(\ell)}(x)S_5^{(\ell)}(x+e_\mu)\rangle_{\beta}^{(\Lambda)} \\
		&=\frac{32g}{|\Lambda|}\sum_{x\in\Lambda}\sum_{\mu=1}^3
		\langle \Gamma^{(1)}(x)\Gamma^{(1)}(x+e_\mu)\rangle_{\beta}^{(\Lambda)}.
	\end{align*}
	Consequently, we obtain
	\begin{equation}
		\label{sumcp2}
		\frac{1}{|\Lambda|}\sum_p \mathfrak{c}_p\le \mathcal{C}_0|\kappa|+32g\mathcal{E}^{(\Lambda)},
	\end{equation}
	where we have written
	\begin{equation}
		\label{calE}
		\mathcal{E}^{(\Lambda)}:=\frac{1}{|\Lambda|}\sum_{x\in\Lambda}\sum_{\mu=1}^3
		\langle \Gamma^{(1)}(x)\Gamma^{(1)}(x+e_\mu)\rangle_{\beta}^{(\Lambda)}.
	\end{equation}
	As shown in Appendix~\ref{LboundcalE}, this quantity $\mathcal{E}^{(\Lambda)}$ is positive at sufficiently low temperatures
	in the strong coupling regime $g/|\kappa|\gg 1$.

	The left-hand side of (\ref{GammaGammacos}) is written
	\begin{equation*}
		\frac{1}{3|\Lambda|}\sum_p\langle \hat{\Gamma}_p^{(1)}\hat{\Gamma}_{-p}^{(1)}\rangle_{\beta}^{(\Lambda)}
		\biggl[\sum_{\mu=1}^3\cos p^{(\mu)}\biggr]
		=\frac{1}{3|\Lambda|}\sum_{x\in\Lambda}\sum_{\mu=1}^3 \langle \Gamma^{(1)}(x)\Gamma^{(1)}(x+e_\mu)\rangle_{\beta}^{(\Lambda)}
		=\frac{1}{3}\mathcal{E}^{(\Lambda)}
	\end{equation*}
	{from} (\ref{hatGammap1}).
	By combining this, (\ref{GammaGammacos}), (\ref{sumcp}) and (\ref{sumcp2}), we obtain
	\begin{align*}
		\frac{1}{3}\mathcal{E}^{(\Lambda)}&\le \delta(\beta)+\frac{\sqrt{3}I_3}{12\sqrt{g}}
		\sqrt{\mathcal{C}_0|\kappa|+32g\mathcal{E}^{(\Lambda)}}+\Sigma_{\rm LRO}^{(\Lambda)}  \\
		&\le\delta(\beta)+\frac{\sqrt{3}I_3}{12}\sqrt{\frac{\mathcal{C}_0|\kappa|}{g}}
		+\frac{\sqrt{6}}{3}I_3\sqrt{\mathcal{E}^{(\Lambda)}}+\Sigma_{\rm LRO}^{(\Lambda)}.
	\end{align*}
	Therefore, the lower bound for the long-range order $\Sigma_{\rm LRO}^{(\Lambda)}$ is given by
	\begin{equation*}
		\frac{1}{3}\sqrt{\mathcal{E}^{(\Lambda)}}\biggl[\sqrt{\mathcal{E}^{(\Lambda)}}-{\sqrt{6}}I_3\biggr]
		-\delta(\beta)-\frac{\sqrt{3}I_3}{12}\sqrt{\frac{\mathcal{C}_0|\kappa|}{g}}\le \Sigma_{\rm LRO}^{(\Lambda)}.
	\end{equation*}
	The lower bound for $\mathcal{E}^{(\Lambda)}$ is given by
	\begin{equation*}
		\mathcal{E}^{(\Lambda)}\ge 6-\frac{3|\kappa|}{g}-\frac{1}{\beta g}\log 2
	\end{equation*}
	as shown in Appendix~\ref{LboundcalE}. Since $I_3=0.68\cdots$, these imply the existence of the long-range order
	at sufficiently low temperatures in the strong coupling regime $|\kappa|/g\ll 1$.

	\appendix

	\section{Chiral rotational symmetry}
	\label{ChiralRot}

	One can check that, for any two $4 \times 4$ matrices, $\mathcal{M}$ and $\mathcal{M}'$, the following relation holds:
	\begin{equation}
		\label{commuPsiMPsiMp}
		[\Psi^\dagger(x)\mathcal{M}\Psi(x),\Psi^\dagger(x)\mathcal{M}'\Psi(x)]
		=\Psi^\dagger(x)[\mathcal{M},\mathcal{M}']\Psi(x).
	\end{equation}
	We write
	\begin{equation}
		\label{Gamma5}
		\Gamma^{(5)}(x):=\Psi^\dagger(x)\gamma_5\Psi(x).
	\end{equation}
	Then, by using the above formula, one has the commutation relations,
	\begin{equation}
		\label{commuGamma}
		\begin{split}
			& [\Gamma^{(1)}(x),\Gamma^{(2)}(x)]=2i\Gamma^{(5)}(x), \quad [\Gamma^{(2)}(x),\Gamma^{(5)}(x)]=2i\Gamma^{(1)}(x), \\
			\mbox{and} \quad & [\Gamma^{(5)}(x),\Gamma^{(1)}(x)]=2i\Gamma^{(2)}(x).
		\end{split}
	\end{equation}
	In addition, {from} the definitions of $S^{(j)}(x)$ for $j=1,2,3$, we have
	\begin{equation}
		\label{commuGamma1S12}
		[\Gamma^{(1)}(x),S^{(j)}(x)]=0 \quad \mbox{for \ } j=1,2,3.
	\end{equation}
	Moreover, one has
	\begin{equation}
		\label{commuGammaS5}
		[\Gamma^{(1)}(x),S_5^{(\ell)}(x)]=2i\Psi^\dagger(x)\gamma_5\tau_\ell\Psi(x)
	\end{equation}
	and
	\begin{equation}
		\label{commuGammaS52}
		[\Gamma^{(1)}(x),\Psi^\dagger(x)\gamma_5\tau_\ell\Psi(x)]=-2iS_5^{(\ell)}(x).
	\end{equation}
	for $\ell=1,2,3$.

	Next consider the rotation by the generator $\Gamma^{(5)}(x)$. Let $\theta\in\mathbb{R}$, and note that
	\begin{align*}
		\frac{d}{d\theta}e^{i\theta\Gamma^{(5)}(x)}\Gamma^{(1)}(x)e^{-i\theta\Gamma^{(5)}(x)}
		&=ie^{i\theta\Gamma^{(5)}(x)}[\Gamma^{(5)}(x),\Gamma^{(1)}(x)]e^{-i\theta\Gamma^{(5)}(x)} \\
		&=-2e^{i\theta\Gamma^{(5)}(x)}\Gamma^{(2)}(x)e^{-i\theta\Gamma^{(5)}(x)},
	\end{align*}
	where we have used the above commutation relation $[\Gamma^{(5)}(x),\Gamma^{(1)}(x)]=2i\Gamma^{(2)}(x)$.
	Further,
	\begin{align*}
		\frac{d^2}{d\theta^2}e^{i\theta\Gamma^{(5)}(x)}\Gamma^{(1)}(x)e^{-i\theta\Gamma^{(5)}(x)}
		&=-2ie^{i\theta\Gamma^{(5)}(x)}[\Gamma^{(5)}(x),\Gamma^{(2)}(x)]e^{-i\theta\Gamma^{(5)}(x)} \\
		&=-4e^{i\theta\Gamma^{(5)}(x)}\Gamma^{(1)}(x)e^{-i\theta\Gamma^{(5)}(x)}.
	\end{align*}
	These imply
	\begin{equation*}
		e^{i\theta\Gamma^{(5)}(x)}\Gamma^{(1)}(x)e^{-i\theta\Gamma^{(5)}(x)}
		= \Gamma^{(1)}(x)\cos 2\theta - \Gamma^{(2)}(x)\sin 2\theta.
	\end{equation*}
	In particular, for $\theta=-\pi/4$, the following holds:
	\begin{equation}
		\label{relatG1G2}
		e^{-i(\pi/4)\Gamma^{(5)}(x)}\Gamma^{(1)}(x)e^{i(\pi/4)\Gamma^{(5)}(x)}= \Gamma^{(2)}(x).
	\end{equation}

	\section{A lower bound for $\mathcal{E}^{(\Lambda)}$}
	\label{LboundcalE}

	In this Appendix, we obtain a lower bound for $\mathcal{E}^{(\Lambda)}$ of (\ref{calE}).

	In the same way as in \cite{GK2}, we first use Peierls's inequality. Namely, we have
	\begin{equation}
		\label{Peierlsineq}
		Z_{\beta}^{(\Lambda)}:={\rm Tr}\exp[-\beta H^{(\Lambda)}(0)]\ge
		\exp[-\beta\langle \Omega_{\rm var}^{(\Lambda)},H^{(\Lambda)}(0)\Omega_{\rm var}^{(\Lambda)}\rangle]
	\end{equation}
	for any variational wavefunction $\Omega_{\rm var}^{(\Lambda)}$. We recall the expression of the operator $\Gamma^{(1)}(x)$, i.e.,
	\begin{equation*}
		\Gamma^{(1)}(x)=\Psi_{\rm u}^\dagger(x)\gamma_0\Psi_{\rm u}(x)+\Psi_{\rm d}^\dagger(x)\gamma_0\Psi_{\rm d}(x).
	\end{equation*}
	Since the matrix $\gamma_0$ in $\Gamma^{(1)}(x)$ has
	the four eigenvalues, $+1,+1,-1,-1$, we choose the variational state $\Omega_{\rm var}^{(\Lambda)}$ so that
	it is the eigenstate of $\Gamma^{(1)}(x)$ for both of the two flavors, u and d,
	and all the sites $x\in\Lambda$ with the eigenvalue $+4$, i.e.,
	$\Gamma^{(1)}(x)\Omega_{\rm var}^{(\Lambda)}=4\Omega_{\rm var}^{(\Lambda)}$.
	In other words, each site $x$ is occupied by the two eigenstates of $\gamma_0$ with the eigenvalue $+1$
	for both of the two flavors, u and d. Then, clearly, one has
	\begin{equation*}
		\langle \Omega_{\rm var}^{(\Lambda)},H_{\rm K}^{(\Lambda)}\Omega_{\rm var}^{(\Lambda)}\rangle=0
	\end{equation*}
	for the kinetic Hamiltonian $H_{\rm K}^{(\Lambda)}$. Further, we have
	\begin{equation*}
		\langle\Omega_{\rm var}^{(\Lambda)},\Gamma^{(2)}(x)\Omega_{\rm var}^{(\Lambda)}\rangle=0
	\end{equation*}
	for all $x\in\Lambda$ because the matrix $\gamma_0\gamma_5$ in $\Gamma^{(2)}(x)$ satisfies
	the anticommutation relation $\gamma_0\gamma_0\gamma_5+\gamma_0\gamma_5\gamma_0=0$, which yields $\langle v,\gamma_0\gamma_5 v\rangle
	+\langle v,\gamma_0\gamma_5 v\rangle=2\langle v,\gamma_0\gamma_5 v\rangle=0$ for any eigenvector $v$ of $\gamma_0$.
	Therefore, except for the term having the operator $\Gamma^{(1)}(x)$, all the expectation values of the terms in the Hamiltonian
	are vanishing. From these observations, one has
	\begin{equation*}
		\langle \Omega_{\rm var}^{(\Lambda)},H^{(\Lambda)}(0)\Omega_{\rm var}^{(\Lambda)}\rangle=-16\cdot 3 g|\Lambda|.
	\end{equation*}
	By substituting this into the right-hand side of (\ref{Peierlsineq}), we have
	\begin{equation}
		\label{Zlowbound}
		\log Z_{\beta}^{(\Lambda)}\ge 16\cdot 3 \beta g|\Lambda|.
	\end{equation}

	On the other hand, the following holds:
	\begin{equation*}
		\log Z_{\beta}^{(\Lambda)}=-\beta\langle H^{(\Lambda)}(0)\rangle_{\beta}^{(\Lambda)}+
		S_{\beta}^{(\Lambda)}\le -\beta\langle H^{(\Lambda)}(0)\rangle_{\beta}^{(\Lambda)}
		+\log 2^{8|\Lambda|},
	\end{equation*}
	where $S_{\beta}^{(\Lambda)}$ is the entropy, and we have used its upper bound.
	By combining this with the above inequality (\ref{Zlowbound}), one has
	\begin{equation*}
		-\frac{1}{|\Lambda|}\langle H^{(\Lambda)}(0)\rangle_{\beta}^{(\Lambda)}\ge
		16\cdot 3  g-\frac{8}{\beta}\log 2.
	\end{equation*}
	The left-hand side is written
	\begin{align*}
		-\frac{1}{|\Lambda|}\langle H^{(\Lambda)}(0)\rangle_{\beta}^{(\Lambda)}
		&=-\frac{1}{|\Lambda|}\langle H_{\rm K}^{(\Lambda)}\rangle_{\beta}^{(\Lambda)}
		+\frac{8g}{|\Lambda|}\sum_{x\in\Lambda}\sum_{\mu=1}^3\langle \Gamma^{(1)}(x)\Gamma^{(1)}(x+e_\mu)\rangle_{\beta}^{(\Lambda)} \\
		&\le8\cdot 3|\kappa|+8g\mathcal{E}^{(\Lambda)},
	\end{align*}
	where we have used the rotational symmetries of the Hamiltonian $H^{(\Lambda)}(0)$.
	Therefore, we obtain the desired lower bound,
	\begin{equation}
		\label{calElowbound}
		\mathcal{E}^{(\Lambda)}\ge 6-\frac{3|\kappa|}{g}-\frac{1}{\beta g}\log 2.
	\end{equation}

	\section{Effective Hamiltonian}
	\label{Sec.EH}

	As mentioned in Sec.~\ref{Sec:Intro}, a reader might doubt the possibility
	that there appear two spontaneous magnetizations with different directions
	in a single phase. In this Appendix, we give a plausible argument
	that two such spontaneous magnetizations are possible to emerge in different directions.

	For this purpose, we introduce an effective massive model \cite{Braghin1,Braghin2}.
	Since a single spontaneous magnetization is already proved to appear
	in our mathematical argument,
	we replace the	effect of one of the two spontaneous magnetizations by the corresponding effective mass term in the Hamiltonian.
	More precisely, the effective Hamiltonian is given by

	\begin{equation*}
		H_{\rm EM}^{(\Lambda)} = H_{\rm K}^{(\Lambda)} + H_{\rm M}^{(\Lambda)} + H_{\rm EI}^{(\Lambda)},
	\end{equation*}
	where the first term is the same kinetic Hamiltonian given by
	\begin{align*}
		H_{\rm K}^{(\Lambda)} := i\kappa \sum_{x \in \Lambda \subset \mathbb{Z}^3} &\big[ \Psi^{\dagger}(x)\alpha_1\Psi(x + e_1)
		- \Psi^{\dagger}(x + e_1)\alpha_1\Psi(x) \nonumber \\
		&+ \Psi^{\dagger}(x)\alpha_2\Psi(x + e_2) - \Psi^{\dagger}(x + e_2)\alpha_2\Psi(x) \\
		&+ \Psi^{\dagger}(x)\alpha_3\Psi(x + e_3) - \Psi^{\dagger}(x + e_3)\alpha_3\Psi(x) \big],
	\end{align*}
	with the hopping parameter $\kappa \in \mathbb{R}$, the second term is the effective mass term which is given by
	\begin{equation*}
		H_{\rm M}^{(\Lambda)} := m \sum_{x \in \Lambda} S^{(1)}(x)
	\end{equation*}
	with the constant $m \in \mathbb{R}$, and the third term is the rest of the interactions,
	\begin{equation*}
		H_{\rm EI}^{(\Lambda)} := -g \sum_{x \in \Lambda} \sum_{\mu=1}^{3} \big[ \Gamma^{(1)}(x)\Gamma^{(1)}(x + e_\mu)
		+ S_5^{(2)}(x)S_5^{(2)}(x + e_\mu) + S_5^{(3)}(x)S_5^{(3)}(x + e_\mu) \big],
	\end{equation*}
	with the coupling constant $g > 0$. In order to see the symmetries of the Hamiltonian $H_{\rm EM}^{(\Lambda)}$, we write
	\begin{equation*}
		\Psi \to e^{i\theta \mathcal{Q}} \Psi
	\end{equation*}
	with a generator $\mathcal{Q}$ and the angle $\theta \in \mathbb{R}$.
	Then, one can check that the Hamiltonian
	$H_{\rm EM}^{(\Lambda)}$ is invariant under the transformations with the generators, $\tau_1, \gamma_5\tau_2, \gamma_5\tau_3$. In addition, one notices that the following commutativity relations hold:
	\begin{equation*}
		[S^{(1)}(x), \Gamma^{(1)}(x)] = 0, \quad [S^{(1)}(x), S_5^{(2)}(x)] = 0, \quad [S^{(1)}(x), S_5^{(3)}(x)] = 0
	\end{equation*}
	for any $x \in \Lambda$. These relations imply that, when there appears a long-range order which is induced by the interaction Hamiltonian $H_{\rm EI}^{(\Lambda)}$, it is expected that the effective mass Hamiltonian $ H_{\rm M}^{(\Lambda)}$ never suppresses the long-range order. Actually, we can prove the existence of the long-range order for the order parameter,
	\begin{equation*}
		O_{5,\Lambda}^{(2)} = \frac{1}{|\Lambda|} \sum_{x \in \Lambda} S_5^{(2)}(x),
	\end{equation*}
	for sufficiently small $|m/g|$ and $|\kappa/g|$ in the same way as in the proof of our main theorem.
	This yields the spontaneous magnetization for the order parameter $O_{5,\Lambda}^{(2)}$.
	Since the mass term is introduced into the Hamiltonian, the model indeed shows the violation of the parity in a pure phase.

	\section{Reflection positivity for the effective mass term}
	\label{Sec.RPEM}
	In this Appendix, we prove that the reflection positivity holds for the effective mass term,
	\begin{equation*}
		H_{\rm M}^{(\Lambda)} := m \sum_{x \in \Lambda} S^{(1)}(x).
	\end{equation*}
	First, in the same way as in the case of the original Hamiltonian, we decompose
	this into two parts as follows:
	\begin{equation*}
		H_{\rm M}^{(\Lambda)} = H_{\rm M}^{+} + H_{\rm M}^{-},
	\end{equation*}
	where
	\begin{equation*}
		H_{\rm M}^{\pm} := m \sum_{x \in \Lambda^{\pm}} S^{(1)}(x).
	\end{equation*}
	Further, we write
	\begin{equation*}
		\tilde{H}_{M}^{\pm} := [U(\gamma_0) U_2]^{\dagger} H_{\rm M}^{\pm} U(\gamma_0) U_2,
	\end{equation*}
	and
	\begin{equation*}
		\hat{H}_{M}^{\pm} := \tilde{U}_1^{\dagger} \tilde{H}_{M}^{\pm} \tilde{U}_1.
	\end{equation*}
	Our aim is to prove
	\begin{equation*}
		\vartheta(\hat{H}_{M}^{-}) = \hat{H}_{M}^{+}.
	\end{equation*}

	From the definitions of the unitary transformations, $U(\gamma_0)$ and $U_2$, one has
	\begin{equation*}
		\tilde S^{(1)}(x) = [U(\gamma_0) U_2]^{\dagger} S^{(1)}(x) U(\gamma_0) U_2 = \Psi^{\dagger}(x) \tau_1 \Psi(x).
	\end{equation*}
	Further, from the definition of the unitary transformation $\tilde{U}_1$, we have
	\begin{equation*}
		\hat S^{(1)}(x) =\begin{cases}
			\mp [\T\Psi_{\rm u}(x) \alpha_1 \Psi_{\rm d}(x) + \Psi_{\rm d}^{\dagger}(x) \alpha_1 \T\Psi_{\rm u}^{\dagger}(x)] \text{ for }x\in \Lambda_{\rm odd}\cap \Lambda_{\pm}\\
			\mp [\Psi_{\rm u}^\dagger(x) \alpha_1 \T\Psi_{\rm d}^\dagger(x)+\T\Psi_{\rm d}(x) \alpha_1 \Psi_{\rm u}(x)] \text{ for } x\in \Lambda_{\rm even}\cap \Lambda_{\pm}.
		\end{cases}
	\end{equation*}
	Since the matrix $\alpha_1$ is real symmetric, this can be written as
	\begin{equation*}
		\hat S^{(1)}(x) =\begin{cases}
			\mp [\T\Psi_{\rm u}(x) \alpha_1 \Psi_{\rm d}(x) + \Psi_{\rm d}^{\dagger}(x) \alpha_1 \T\Psi_{\rm u}^{\dagger}(x)] \text{ for }x\in \Lambda_{\rm odd}\cap \Lambda_{\pm}\\
			\pm [ \T\Psi_{\rm u}(x) \alpha_1 \Psi_{\rm d}(x)+\Psi_{\rm d}^\dagger(x) \alpha_1 \T\Psi_{\rm u}^\dagger(x)] \text{ for } x\in \Lambda_{\rm even}\cap \Lambda_{\pm}.
		\end{cases}
	\end{equation*}
	by using the anticommutation relations for the fermion operators.
	This implies that the reflection of $\hat S^{(1)}(x)$ does not change its sign. Thus, the relation holds.

	\bigskip

	\subsection*{Acknowledgements}
	We would like to thank Sinya Aoki for valuable discussions.
	The present paper is an answer to his question whether the methods developed in \cite{GK,GK2} for
	the Nambu--Jona-Lasinio models are applicable to the mathematical proof of the parity violation which is conjectured in the papers
	\cite{ABG,AG}.
	We are also grateful to Etsuko Itou for her helpful comments on the presentation of the paper.
	Y.~G. is partially supported by JSPS Kakenhi Grant Number 23K12989.


\end{document}